\documentclass[prl,aps,twocolumn,groupedaddress,showpacs,floatfix,preprintnumbers,superscriptaddress]{revtex4-1}
\usepackage{amsmath,amssymb,multirow,bm}
\usepackage{soul} 
\usepackage{xcolor}
\usepackage{citesort}

\newcommand{\beq}{\begin{equation}}
\newcommand{\eeq}{\end{equation}}
\newcommand{\bea}{\begin{eqnarray}}
\newcommand{\eea}{\end{eqnarray}}
\usepackage{graphicx}
\preprint{}

\begin{document}

\title{Parton distribution functions probed in ultraperipheral collisions at the CERN Large Hadron Collider}
\author{J. Thomas}
\email{jamesothomas@gmail.com}
\affiliation{Department of Physics and Astronomy, Texas A\&M University-Commerce, Commerce, Texas 75429, USA}
\author{C.A. Bertulani}
\email{carlos.bertulani@tamuc.edu}
\affiliation{Department of Physics and Astronomy, Texas A\&M University-Commerce, Commerce, Texas 75429, USA}
\affiliation{Department of Physics and Astronomy, Texas A\&M University, College Station, Texas 77843, USA}
\author{N. Brady}
\email{nbrady@leomail.tamuc.edu}
\affiliation{Department of Physics and Astronomy, Texas A\&M University-Commerce, Commerce, Texas 75429, USA}
\author{D. B. Clark}
\email{dbclark@mail.smu.edu}
\affiliation{Department of Physics, Southern Methodist University, Dallas, Texas 75205, USA}
\author{E. Godat}
\email{egodat@mail.smu.edu}
\affiliation{Department of Physics, Southern Methodist University, Dallas, Texas 75205, USA}
\author{F. Olness}
\email{olness@smu.edu}
\affiliation{Department of Physics, Southern Methodist University, Dallas, Texas 75205, USA}

\date{\today}

\begin{abstract}

Vector meson production in ultra-peripheral pA and AA collisions at the CERN Large Hadron Collider (LHC) are very sensitive to Parton Distribution Functions (PDF) as well as  to their leading-order, next-to-leading-order, and medium corrections. This process is a complimentary tool to explore the effects of different PDFs in particle production in proton-nucleus and nucleus-nucleus central collisions. Existing and forthcoming data available, e.g., from ALICE and CMS, may be used in conjunction with our theoretical predictions to constrain the PDFs.  We make predictions for rapidity distributions and for cross sections of J/$\psi$ , $\psi(2S)$ and $\Upsilon$ production at $\sqrt{s_{NN}}=2.76$ TeV and $\sqrt{s_{NN}}=5$ TeV.  We use the second energy as representative for the Run 2 of PbPb collisions at the LHC.

\end{abstract}


\maketitle

\section{Introduction}

The Large Hadron Collider (LHC) at CERN has made seminal contributions to particle, nuclear and atomic physics, contributing with numerous scientific discoveries and tests of fundamental theoretical predictions in the physical sciences (see, e.g., Refs. \cite{Zepto,HE13}). Proton-proton (pp) proton-nucleus (pA) and nucleus-nucleus (AA) collisions produced in heavy ion accelerators such as the GSI facility in Germany, the RHIC facility in the U.S. and at the LHC in Switzerland have been looking for particles predicted by the standard model (SM) \cite{Eng64,Hig64,Gur64}, as well as new particles, beyond the SM, such as supersymmetric particles \cite{Miy66,DG81,Mur00}, and for extra-dimensions \cite{Kal21,Kle26,RS99,Ran02}. AA collisions have also been used to probe a phase-transition of nuclear matter, from a hadronic to a quark-gluon phase \cite{CP75,HN77,Zaj08}.  For example, charmonium states, such as the J/$\psi$ particles, are commonly used as probes of the strongly interacting matter occurring in relativistic heavy ion collisions.  Due to their small size ($< 1$ fm) and relatively large binding energy (hundreds of MeV) they are considered ideal probes of the phase transtion to a quark-gluon plasma (QGP). The production of charmonium states has been predicted \cite{Mat86} to be suppressed and is considered a good signature of the  QGP.  This has indeed been observed  at the RHIC accelerator at Brookhaven, and at the CERN SPS and LHC \cite{Al05,Abe09,Amal07,ATL11,Ada11,ALI12,CMS12}.

Hadron accelerators at CERN and elsewhere have also been used for studies of photo-nuclear and photon-photon physics induced in ultraperipheral heavy ion collisions (UPC), i.e., relativistic collisions without nuclear contact \cite{BB94}. It has long been realized  that the strong electromagnetic fields produced during a short time in such accelerators can used to probe new phenomena in atomic, nuclear and particle physics \cite{BB88}. For example, the double giant dipole resonance (DGDR), a giant resonance (GR) excited on top of another GR was produced at the GSI facility in 1993 by means of UPCs with heavy ions \cite{Rit93,Sch93}. In 1996, the first production of anti-hydrogen atoms in the laboratory was carried out using UPCs in CERN \cite{Bau96}, followed by another experiment in Fermilab \cite{Bla98}. More recently, UPCs have been used to study gluon distributions in nuclei via the production of vector mesons, such as  J/$\psi$  production in recent experiments at the LHC \cite{GB02,Fran02,Gut13,AB11,AB12,AN13,Reb12,CMS14,Abb13,Abe13,Abe14,Ada15,GM05,GM06,SS07,RS08,AG08,Bert09,CE10,GM11,CS12,GM12,GM14,GMN14,GMN15}. This opened another opportunity to study  the manifestation of sub-nucleon degrees of freedom not only in dense QGP matter, but also in interactions of quasi-real photons with partonic distributions within the nuclei. Such studies are complementary and greatly enhance our understanding of the strong interaction and of QCD in general. 

The basic idea of UPCs is that the strong electromagnetic fields of a relativistic heavy ion due to the Lorentz contraction  give rise to hard (high energy) photons that will further react with the proton or nucleus moving in the opposite direction in the collider. Such peripheral collisions give rise to numerous physical phenomena of interest for nuclear and particle physics such as the study of radiative meson widths, photo-production of exotic mesons, and ultimately the tests of parton distribution functions. Many theoretical publications on UPCs have appeared in the last two decades, followed by a good number of experiments at the GSI, RHIC and the LHC. An incomplete list of related works is listed in Refs. \cite{BB94,BB88,Rit93,Sch93,Bau96,Bla98,GB02,Fran02,Gut13,AB11,AB12,AN13,Reb12,CMS14,Abb13,Abe13,Abe14,Ada15,GM05,GM06,SS07,RS08,AG08,Bert09,CE10,GM11,CS12,GM12,GM14,GMN14,GMN15,Bau02,BKN05,Bal08} 

 A recent study   \cite{And15} has shown that the experimental data on J/$\psi$ production in ultraperipheral collisions at the LHC cannot be reproduced by means of production, rescattering, and destruction of  J/$\psi$'s with hadrons alone, without the account of subnucleonic degrees of freedom. UPC data are good probes of PDFs  at small momentum fraction $x$ and momentum transfers $Q^2$ up to a few GeV$^2$ for J/$\psi$ production and up to a few tens of GeV$^2$ for $\Upsilon$ production. In this article we explore the details of the rapidity distributions and total cross sections for J/$\psi$, $\psi$(2S)  and $\Upsilon$ production at the LHC using subnucleonic degres of freedom as with some of the previous published works, e.g., Refs. \cite{GB02,AB11,AB12}.  Our goal is to test how several PDFs fare against experimental verification. In this work we have extended the study performed in Refs. \cite{AB11,AB12,ABM12} to account for tests of parton distribution functions at the LHC, also including other  PDFs and making predictions for  the J/$\psi$, $\Upsilon(1s)$, and  $\psi(2S)$.

In section 2 we make a short review of the reaction mechanism adopted for the description of particle production in UPCs. In section 3 we summarize the main physics in each of the PDFs adopted in this work. In section 4 we present our experimental results, comparing them with the few existing experimental data and making predictions for future experiments. Our conclusions are spelled out in section 5.  

\section{Reaction mechanism}

Peripheral heavy ion collisions have been reviewed in previous publications \cite{BB88,Bau02,BKN05}. In particular, the appropriate reaction mechanism to use in UPC as a probe of gluon distribution functions has been described in Ref. \cite{GB02} and used as a standard method in other publications (e.g., see Refs. \cite{AB11,AB12}). Here we summarize the main equations that have been used in our calculations with some small modifications from previous works.   

The cross section for the production of a vector meson $V$ in UPC between heavy ions (AA collisions) is given by
\beq
\sigma^{AA \rightarrow V}=2 \int d\omega \,\frac{N(\omega)}{\omega}\, \sigma^{\gamma A
  \rightarrow V}(\omega) \, ,
\label{sigAA}
\eeq
where $\omega$ is the (virtual) photon energy, $\sigma^{\gamma A  \rightarrow V}(\omega)$ is the photoproduction cross section and $N(\omega)$ are the virtual photon numbers \cite{BB88}.  The factor of $2$  accounts for the production of mesons in the target and in the projectile, which are equal in magnitude.  At a given collision impact parameter $b$ the virtual photon flux is given by (we use $\hbar=c=1$)
\beq
 n(\omega, b) = \dfrac{Z^2\alpha}{\pi^2}\dfrac{\xi^2}{b^2} \left[ K_1^2(\xi) + \dfrac{1}{\gamma^2}K_0^2(\xi) \right] e^{-T(b)},
 \label{nob}
\eeq
with $\xi = {\omega b}/\gamma $, 
where  $Z$ is the nuclear charge,  $\gamma$ is the Lorentz contraction factor, and $K_0$  and $K_1$  are modified Bessel functions. The exponential factor accounts for the probability that both ions survive in the collision, with the profile function $T(b)$ given by
\beq
 T(b) = \dfrac{\sigma_{NN}}{2\pi}\int_0^\infty dq\, q\, \left[\tilde{\rho_A}(q)\right]^2  J_0(qb), 
 \label{Tb}
\eeq
where $\sigma_{NN}$ is the total nucleon-nucleon cross section,  $\tilde{\rho}_{A}(q)$ is the Fourier transform of the nuclear matter density (assumed to be spherical), and $J_0$ is the cylindrical Bessel function.  For $pA \rightarrow V$  processes one just replaces  $\left[\tilde{\rho_A}(q)\right]^2 \longrightarrow \tilde{\rho_p}(q) \tilde{\rho_A}(q)$ in Eq. \eqref{Tb}, where $\tilde{\rho_p}(q)$ now is the proton form factor.

The virtual photon numbers entering Eq. \ref{sigAA} are obtained by the integral of $n(\omega, b) $ over all impact parameters, also accounting for photon polarization effects (for more details, see Refs. \cite{BFF90,CJ90}),  
\beq
N(\omega) =   2 \pi \int db \, b  \int {dr \, r \over C_A}  \int_0^{2\pi} d\phi \ n(\omega,b+r\cos \phi)\, , \label{nomeg} 
\eeq  
where $C_A$ is the cross sectional area of the nucleus which can be obtained from its matter density distribution as a function of the intrinsic coordinate, $\rho_A(r)$. In this equation we have introduced a modification of the method used in Refs. \cite{BFF90,CJ90} which treats nuclei as hard spheres, and we introduce a proper description of the matter diffusiveness in the nucleus. This is important because the most effective collisions for vector meson production occur when the nuclei fly by very close to each other, i.e., in ``grazing" collisions. In our method there is no need to introduce a cutoff impact parameter. The meson is expected to be produced with transverse polarization due to helicity conservation of a quasi-real photon \cite{BFF90,CJ90}. 

For the rapidity distributions, $d\sigma^{A(p)A \rightarrow V}/dy$, we use the formalism of Refs. \cite{GB02,AB11}, so that the kinematical relation $d\sigma/dy = \omega d\sigma/d\omega$ is used and the corresponding distributions become 
\bea
\frac{d \sigma^{pA \rightarrow V}}{dy}&=& 
\bigg[ N^A(\omega) \sigma^{\gamma p
\rightarrow V}(\omega)\bigg]_{\omega=\omega_l} \nonumber \\
&+&\bigg[N^p(\omega) \sigma^{\gamma A
\rightarrow V}(\omega)\bigg]_{\omega=\omega_r} 
\label{dsigpA}
\eea
where $\omega_l$ ($\omega_l \propto e^{-y}$) and $\omega_r$ ($\omega_r \propto e^{y}$)  denote photons arising from A and  from p, respectively, with nuclei assumed to come from the left and protons from the right of the observer. This equation takes care of the asymmetry in the number of photons incident on the proton and on the nucleus, respectively. For AA collisions, 
\bea
\frac{d \sigma^{AA \rightarrow V}}{dy}&=& 
\bigg[ N^A(\omega) \sigma^{\gamma A
\rightarrow V}(\omega)\bigg]_{\omega=\omega_l} \nonumber \\
&+&\bigg[N^A(\omega) \sigma^{\gamma A
\rightarrow V}(\omega)\bigg]_{\omega=\omega_r} 
\label{dsigAA}
\eea
and the rapidity distributions are obviously symmetric about mid-rapidity, $y=0$ because the left and right photon fluxes are identical and the second term in this equation yields a mirror image distribution  of the first term as a function of the photon energy in the laboratory, $\omega$. The rapidity $y$ is is given by $y = \ln(2\omega/M_V )= \ln(W^2_{\gamma p}/2\gamma_L m_N M_V )$, where $m_N$ is the nucleon mass,
$W_{\gamma p}=\sqrt{s_{\gamma p}}$ is the invariant c.m. energy of the photon-proton system, and $\gamma_L$ is the laboratory Lorentz factor of  the nucleus.

In Fock space, the UPC produced photon can fluctuate leading to the interaction of either bare photons with the nucleus or of a partonic component of the photon with the nucleus \cite{GRV92,SS95,Cor04}. Thus, the cross section for the production of a vector meson in UPC entering Eq. \eqref{sigAA}  can be written as the sum of resolved and unresolved photon fluctuations, i.e.,
\beq
\sigma^{\gamma A  \rightarrow V}(\omega)  = \sigma^{\gamma A  \rightarrow V}_{unres}(\omega)+ \sigma^{\gamma A   \rightarrow V}_{res}(\omega).
\eeq
To calculate the resolved  photoproduction cross section one needs the partonic distributions of  light quarks and gluons in the photons, and in the nuclei as well \cite{KNV02}. A detailed study of the contribution of resolved photons for the production of free $q\bar q$-pairs in UPCs was done in Refs. \cite{KNV02,AB12}. Although the resolved component is a higher-order process, it was shown in Ref. \cite{AB12} that the contributions of the resolved photon  are not negligible, e.g., at the level of about 10\% for the production of $c\bar c$ pairs  and 20\%  for $b \bar b$ pairs. However, the resolved component contribution to the production of bound states  in UPCs is negligible and therefore we will only consider here the contributions of ``bare" unresolved photons interacting with gluon distributions in the nuclei via photon-gluon fusion.

According to Refs. \cite{Rys93,Brod94,Rys97,FKS98,FMS01,Sus00} the cross sections for photo-production of vector mesons to lowest-order is given by
\bea
\sigma^{\gamma A  \rightarrow V}(\omega) &=&  {16\pi^3 \alpha_s^2 \Gamma_{ee} \over 3 \alpha M_V^5}\left[ xg_p(x,Q^2)R_A(x,Q^2)\right]^2 \nonumber \\ 
&\times&\left\{\begin{array}{cc}
  \int _{t_{min}}^\infty dt \left|F(t)\right|^2    &  {\rm for} \ \gamma A  \ {\rm collisions}\\ 
   1/b   &     {\rm for} \ \gamma p  \ {\rm collisions}
\end{array} \right. ,
\label{gamA}
\eea
where, $\alpha$ is the fine-structure constant, $M_V$ is the vector meson  mass, $x = M^2_V/W^2_{\gamma p}$ is the momentum fraction carried by the gluons (Bjorken variable), $g_p(x,Q^2)$ is the proton gluon distribution at the momentum transfer $\bar{Q^2} = (M_V/2)^2$,  $\Gamma_{ee}$ is the meson leptonic decay width, and $\alpha_s(\bar{Q^2})$ is the strong coupling constant calculated at the average momentum transfer $\bar Q^2$.   The relation between the Bjorken $x$ variable and rapidity is given by $x=(M_V/2\omega_L)e^{-y}$, where $\omega_L$ is the photon energy in the laboratory reference frame. 

The function $R_A (x,Q^2)$ accounts for the nuclear medium effects on the modification of the  gluon distribution. For nuclei, the integral in the last term includes the integration over the momentum transfer of the nuclear form factor $F(t)$ with the minimum momentum transfer given by $t_{min}=(M_V^2 /4\omega \gamma_L)^2$, appropriate for narrow resonances. For photo-production on protons, we use the second option in Eq. \eqref{gamA} with $b=4.5$ GeV$^{-2}$ being the slope parameter of the proton form factor. The formalism described in this paragraph is reminiscent of that introduced in Ref. \cite{GB02}, with the assumption that  the forward scattering amplitude determines the photo-production by means of the optical theorem, whereas the elastic momentum transfer is determined by the nucleus (proton) form factor which contains information on the spatial matter distribution. The  photoproduction of vector mesons in UPCs depends significantly more on the medium effects for PbPb collisions than for pPb collisions. We thus focus our study on PbPb collisions which have been studied experimentally at the LHC at the energies $\sqrt{s_{NN}}=2.76$ TeV  and  $\sqrt{s_{NN}}=5$ TeV.
We notice that the energy 2.76 TeV is the equivalent to the 7 TeV run for pp but with PbPb collisions at the LHC. The 5 TeV run is the pPb run for a proton beam energy of 8 TeV. These  are both Run 1 at the LHC. The equivalent PbPb runs for Run 2 are: (a) for 13 TeV the pp collisions occur at 5.125 TeV and the 14 TeV pp collisions equivalent occurs at 5.52 TeV.  Here we use $\sqrt{s_{NN}}=5$ TeV as representative of PbPb collisions for Run 2 at the LHC.

\section{Parton distribution functions and medium corrections}
The PDFs we use in our calculations are scale dependent functions of momentum fraction for partonic degree of freedom in the initial state hadron. The spin direction of the quarks  is not considered separately and the PDFs quantify the   probability densities  for finding a parton  with momentum fraction $x$ (Bjorken variable) at a momentum transfer $Q^2$ in the hadron (here, the proton or nucleus). Usually, these  densities increase with increasing $Q^2$ at low $x$ and decrease with $Q^2$ at high $x$. For a fixed low value of  $Q^2$ the valence quarks are dominant in the nucleon, whereas at  large $Q^2$ values,  the  $q\bar q$-pairs (sea quarks) increase with a low momentum fraction $x$. These quarks and anti-quarks are responsible for only about half of the nucleon momentum, with the other half of its momentum being carried by the gluons. The gluon  PDF also  increases with increasing $Q^2$ and  is the most relevant part in our calculations (for a review, see, e.g., Ref. \cite{Ji04}).

Parton Distribution Functions result from the application of an appropriate factorization theorem \cite{Collins:1987pm}. Factorization separates 
the long distance low-scale interactions from the short-distance interaction. The latter can be calculated in perturbative Quantum Chromodynamics (pQCD) using the property of   asymptotic freedom \cite{GW73,Pol73}. At high $Q^2$, the strong interaction coupling $\alpha_S(Q^2) \ll 1$ and the interaction can be written as an expansion in this  parameter with the leading order (LO) term proportional to $\alpha_S$, the next-to-leading (NLO) term proportional to $\alpha_S^2$, and so on. This approach,  however, fails to describe the low $Q^2$ dynamics where $\alpha_S(Q^2) \sim 1$. 

The low-scale interactions are non-perturbative objects and cannot be predicted with current techniques. Rather, the PDFs are obtained from 
phenomenological fits to abundant experimental data, mostly Deep Inelastic Scattering (DIS) of leptons on nucleons and nuclei. Typical work in this 
direction has been published in several instances in  global fits such as the CTEQ6 \cite{Pum02,KLO04,Nad08}, MSTW08 \cite{MSTW09}, 
GJR08 \cite{GJR08}, NNPDF2 \cite{Bal10}, ABKM010 \cite{ABKM10}, or HERAPDF1 \cite{Aa10} fits to  data from the HERA and Tevatron colliders.

\begin{figure}
\begin{center}
\includegraphics[scale=0.51]{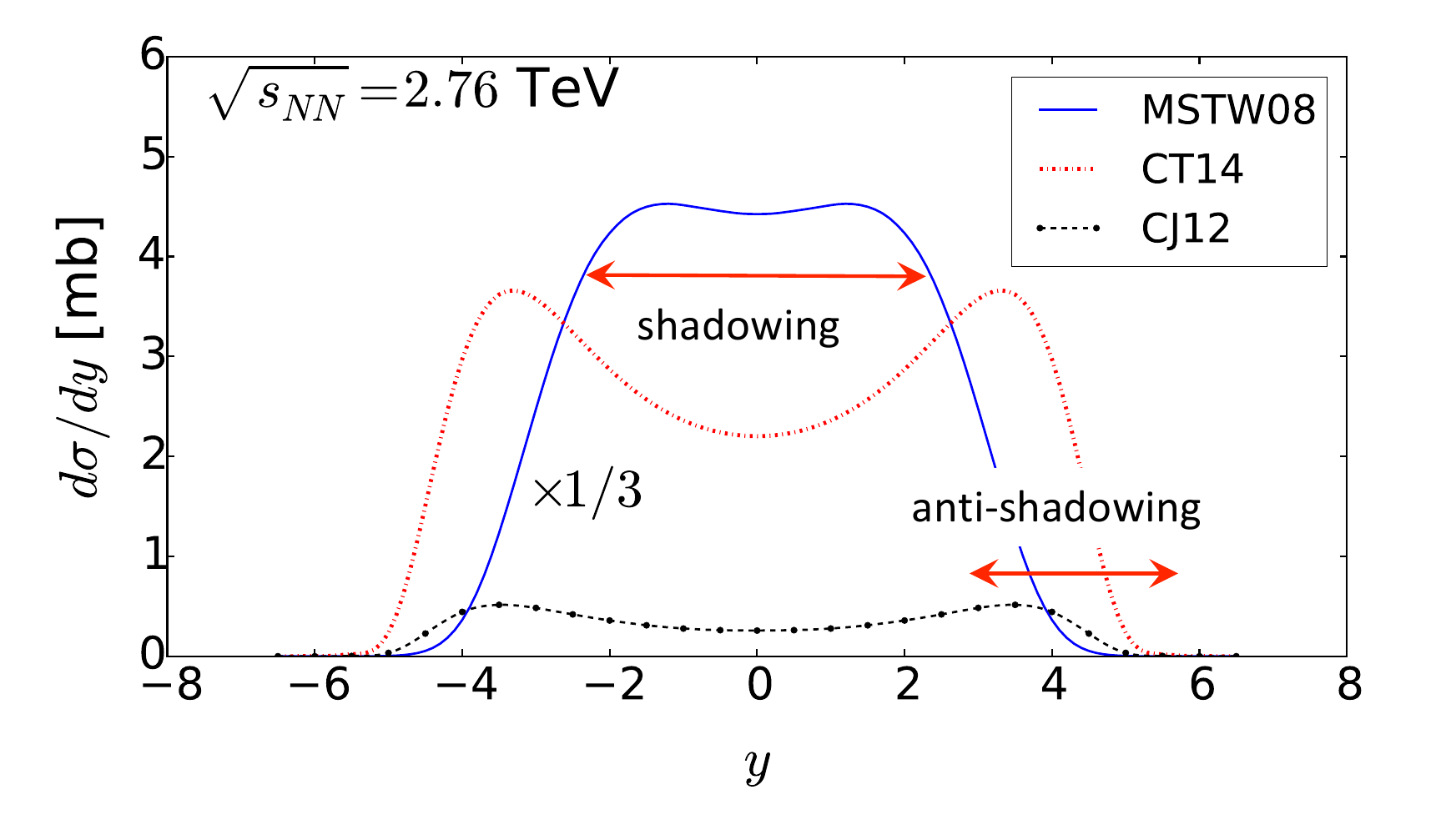}
\includegraphics[scale=0.16]{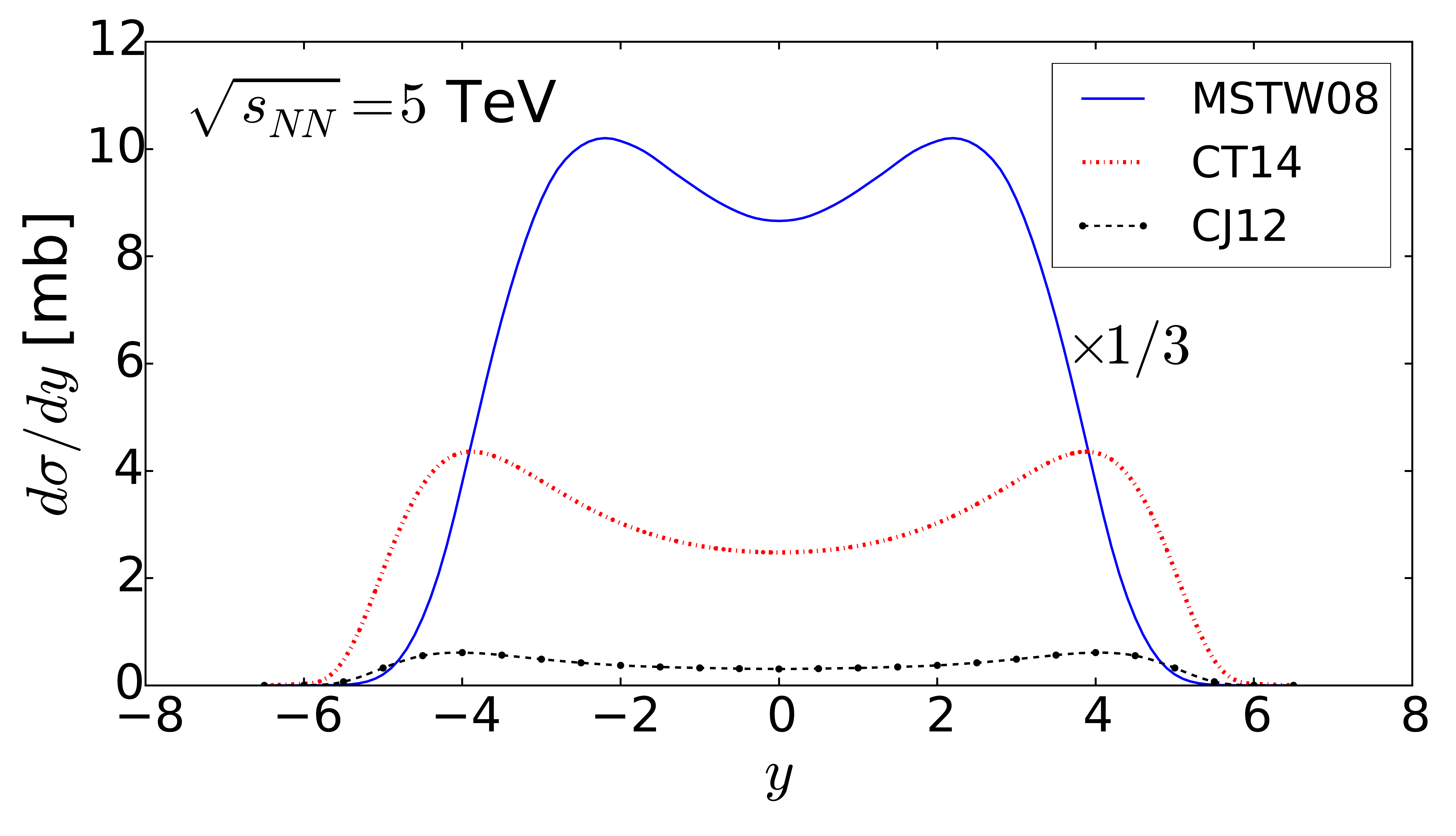}
\caption{Rapidity distributions for J/$\psi$ production in UPCs at the LHC with $\sqrt{s_{NN}}=2.76$ TeV (upper figure) and  $\sqrt{s_{NN}}=5$ TeV (lower figure) using the PDF compilations MSTW08 \cite{MSTW09},  CJ12 \cite{Owe13}, and CT14 \cite{Say14}. The contribution of the MSTW08 set in LO includes a correction factor $\zeta_V$ and its corresponding $d\sigma/dy$ has been multiplied by a factor $1/3$ for better visualization.} \label{f1}
\end{center}
\end{figure}

The largest amount of data used to extract PDFs comes  from deeply inelastic scattering (DIS),  Drell-Yan, and direct jet production  processes. Gluons, being electrically neutral, are not directly  probed in DIS and their distributions are  inferred from the momentum sum rules and the QCD evolution of the PDFs. In the case of nuclei, there are much fewer data  than for the free proton, which makes it interesting to deduce medium effects in nuclei and their dependence on their masses. Since this mass dependence is not well known, medium corrections published in the literature tend to vary  wildly, and effects such as shadowing and antishadowing differ significantly in magnitude \cite{Arm06,EKS99,FS04,HKS04,HKS05,HKN07}. These differences in medium corrections are particularly visible at low $Q^2$. The first global fits adopted mainly DIS and DY lepton pair production and only later other hadronic processes such as particle production in nucleus-nucleus collisions have been included. These global fits have obviously improved the determination of gluon distributions, but large uncertainties at both low and high $x$ are still found in the context of medium corrections and nuclear mass dependence \cite{EPS08,EPS09,FGS05}. The UPC data from the LHC  shows that theoretical calculations using LO and NLO corrections in the PDFs  can provide useful constraints on the small-x dependence of $g_A(x,Q^2)$ \cite{AB11,AB12,ABM12}.

Several PDFs have been used as input in theoretical calculations of meson production in UPCs. For example, in Ref. \cite{GB02} relatively old versions of the PDFs from Refs. \cite{GR78,GRV92,GRV95,EKR98,EKS99,AG01} were used. It was shown that the photo-nuclear vector meson cross sections using the EKS \cite{EKS99} parametrization are about 1.25 larger than that with the AG \cite{AG01} parametrization, while the  GRV  \cite{GRV95} yields a factor 2.4 larger than the EKS prediction.  A similar tendency was found for vector meson production and $c\bar c$ pairs in UPCs in lead-lead collisions at the LHC. Medium effects through the inclusion of the function $R_A(x,Q^2)$ in Eq. \eqref{gamA} were found to be relevant. The same formalism elaborated in Ref. \cite{GB02}  was later adopted in Ref. \cite{Fran02}  to study  color transparency effects in the production of  J/$\psi$ mesons in UPCs. Their calculations yield a bump-shape rapidity distribution with a suppression at mid-rapidity by a factor of about 6 due to the color transparency effect \cite{Fran02}. 

The medium effects included in $R_A(x,Q^2)$ are usually separated into a few regimes: (a) for $x < 0.04$, it is denoted  shadowing, because  the nuclear PDFs are smaller than the  corresponding distributions for free protons, meaning that $R_A < 1$; (b)  antishadowing, occurs as an enhancement ($R_A > 1$) in the range $0.04<x<0.3$; (c)  the EMC effect occurs in the region $0.3 <x<0.8$ with a depletion of the nuclear PDF; and finally (d) for $x>0.8$ one has the region of  Fermi motion with another enhancement  of the nuclear PDFs  \cite{Aub83,GST95,PW00,Arm06}. The medium modification of the  gluon distributions in Pb, using  $R_A (x, Q^2 = M_{J/\psi}^2 /4)$, from the EPS08, EPS09, and HKN07  cases is shown in Figure 1 of Ref. \cite{AB12}, which will be used for the interpretation of some of our  numerical results.

A renewed interest on using UPCs to test PDFs surged with new experimental interest at the LHC. In Ref. \cite{AB11} an investigation of the medium effects encoded in $R_A(x,Q^2)$ was published using the MSTW08 \cite{MSTW09}, EPS08, EPS09 \cite{EPS08,EPS09} and HKN07 \cite{HKN07} PDFs. The parton distributions from Ref.  \cite{MSTW09} (MSTW08) have the advantage of including  free nucleon PDFs to which  nuclear modifications have been included, but they also contain separately nuclear PDFs. In Ref. \cite{AB12,ABM12} these PDFs were studied further with $R_A(x,Q^2)$ calculated  at the factorization scale $Q^2 = (M_{J/\psi}/2)^2$, used in the elastic photoproduction cross sections for the $J/\psi$ meson. For MSTW08  no nuclear effects ($R_A = 1$) are assumed initially, while in HKN07 a weak gluon shadowing prevails with no antishadowing and  EMC effects, but with Fermi motion included. EPS09 has a rather strong shadowing, antishadowing, EMC effect, and Fermi motion effect. But medium effects are strongest in EPS08. These findings have been confirmed by means of our newest calculations presented in this paper. We have introduced some modifications in the way one calculates the number of equivalent photons and also included additional PDFs which can be used as predictions of the higher energy runs at the LHC.

\begin{figure}
\begin{center}
\includegraphics[scale=0.15]{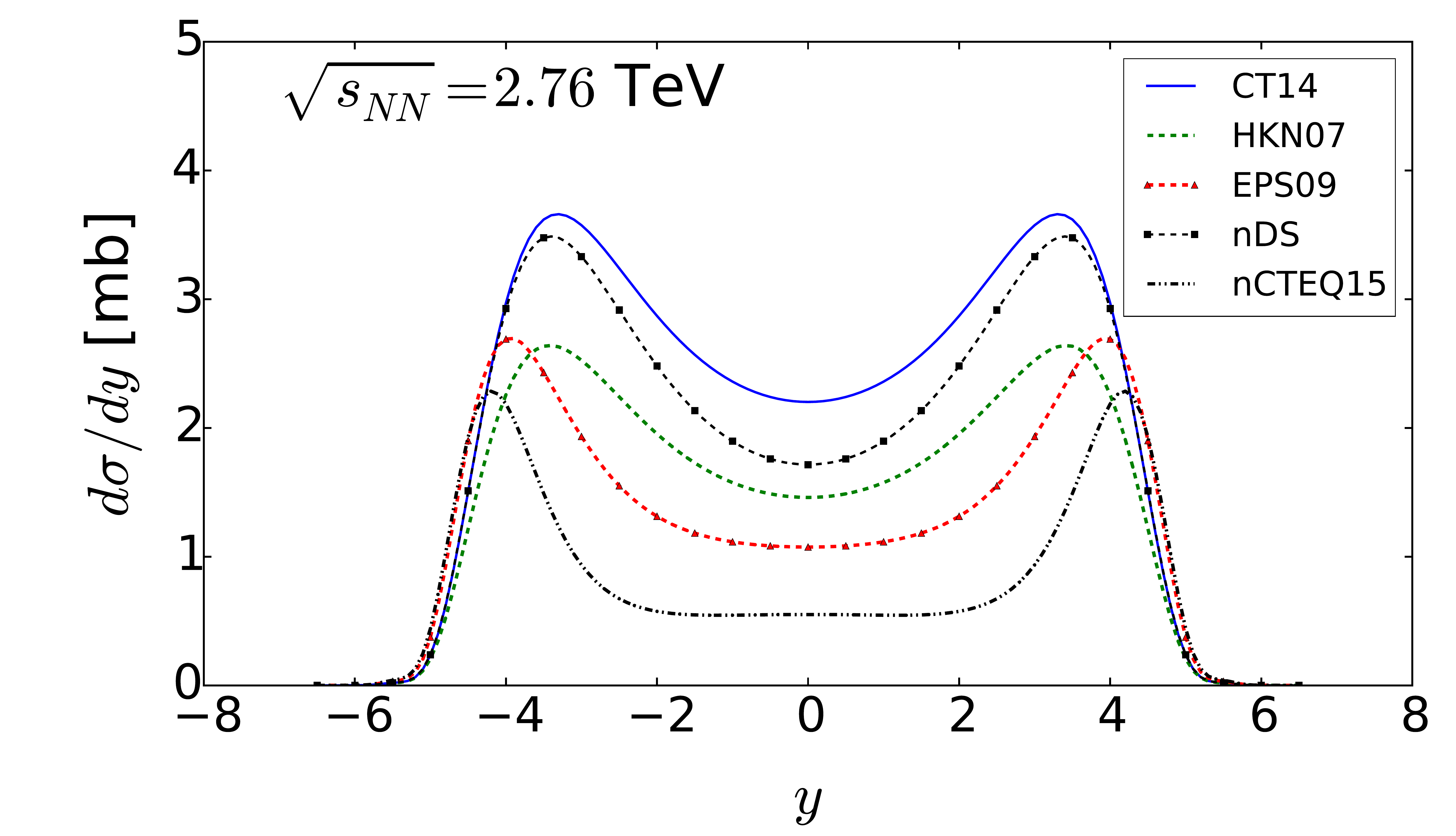}
\includegraphics[scale=0.15]{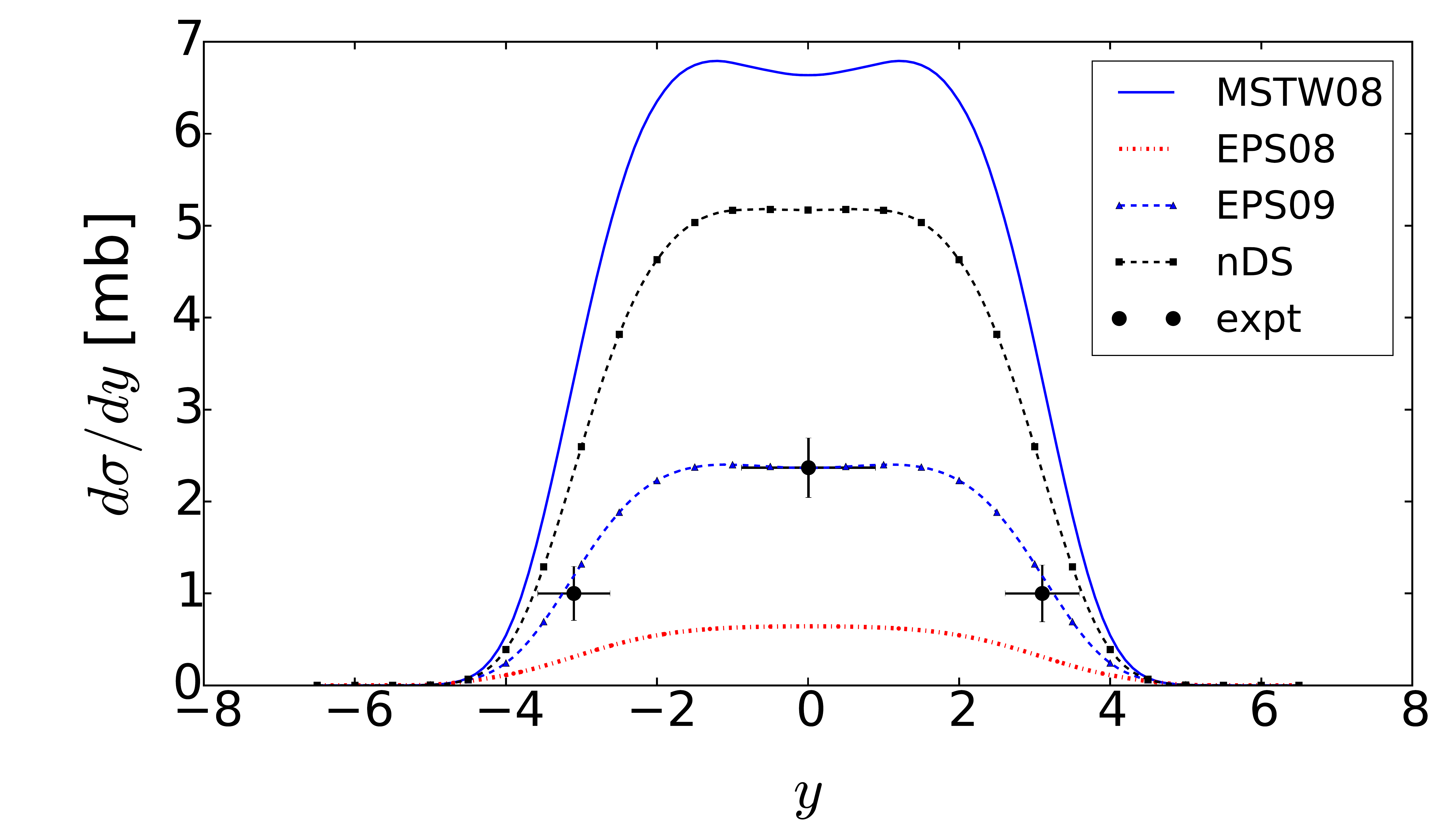}
\caption{Rapidity distributions (in mb) for J/$\psi$ production in UPCs at the LHC with $\sqrt{s_{NN}}=2.76$ TeV   using the EPS09, nDS and HKN07 medium corrections on  CT14 distribution (upper figure). Also shown are the results using nCTEQ15 distributions. The lower figure is the coherent part J/$\psi$ production in UPCs at the LHC using the MSTW08 parton distribution function. The data are from Ref. \cite{Abb13}. } \label{f2}
\end{center}
\end{figure}

\section{Results and discussions}
The results of our calculations for vector meson production of J/$\psi(1S)$ ($M_V=3.096$ GeV), $\Upsilon(1S)$ ($M_V=9.46$ GeV), and  $\psi(2S)$ ($M_V=3.686$ GeV) in UPCs will now be discussed. We choose two energies available at the LHC, $\sqrt{s_{NN}}=2.76$ and 5 TeV. As mentioned earlier, the second energy will be used as representative of the Run 2 for PbPb collisions at the LHC. We start our reporting with J/$\psi$ production, for which experimental data is available  \cite{Abb13,Abe13,Abe14,CMS14,Ada15}. In later sections we will report similar calculations for the corresponding cases of the other two mesons.  While $\psi(2S)$ is also a charmonium state, $\Upsilon(1S)$  is a $b\bar b$ bound state. In all calculations, we have used the nuclear form factor $F(t)$ in Eq. \eqref{gamA} calculated from a Woods-Saxon distribution of the form $\rho(r) = \rho_0/[1 + \exp [(r -R)/a]$, for the nuclear matter,  with $\rho_0$ being the central density, normalized so that $\int \rho(r) d^3r = A$. The nuclear mean radius is given $R$, and the nuclear surface  diffuseness is given by $a$. For $^{208}$Pb the parameters $R = 1.2A^{1/3}$ fm, and $a = 0.549$ fm  reproduce quite well the experimental data of  electron-nucleus scattering \cite{Jag74}.

\subsection{J/$\psi$ production in UPCs at $\sqrt{s_{NN}}=2.76$ TeV and $\sqrt{s_{NN}}=5$ TeV }

We now discuss the rapidity distributions for  J/$\psi$  production at  $\sqrt{s_{NN}}=2.76$ TeV and $\sqrt{s_{NN}}=5$ TeV with PDFs including medium corrections for Pb + Pb collisions.  The J/$\psi$ is the first excited state of $c\bar c$ bound states, usually called ``charmonium". It has the second-smallest rest mass of the charmonia, $J^{PC} = 1^{--}$, a lifetime of $7.2\times 10^{-21}$ s, and a leptonic decay width of $\Gamma_{ee} = 5.55$ keV \cite{PDG}.  For the parton distributions we use the CJ12 \cite{Owe13}, CT14 \cite{Say14}, and the nCTEQ15 \cite{Kov15} releases. The last one already accounts for medium corrections and will be treated without the addition of alternative modifications. The  CJ12 distribution will be modified with medium corrections, $R_A(x,Q^2)$, as defined in Eq.  \eqref{gamA}.  These corrections include the nDS \cite{FS04}, HKN07 \cite{HKN07}  and EPS08, EPS09  \cite{EPS08,EPS09}.  For the last case, we use in fact only  the updated version, i.e., EPS09, except in one case of illustrative value (bottom panel of Figure \ref{f2}).  

Initially, we do not include medium corrections for the CJ12 \cite{Owe13}, and CT14 \cite{Say14} releases. Our results for $\sqrt{s_{NN}}=2.76$ TeV  are shown in the upper panel of Figure \ref{f1}. For the MSTW08  \cite{MSTW09} we use the leading order (LO) and include Next-to-Leading-Order (NLO) corrections in the meson photo-production cross section  through a phenomenological multiplicative factor $\zeta_V=1/3.5$. As explained in  Ref. \cite{AB11}, this factor is found by adjusting the meson photoproduction  data on protons calculated theoretically, $\sigma^{\gamma p \rightarrow V}$, with experimental data from HERA. The excellent agreement with the experimental data is attested in Figure 2 of  Ref. \cite{AB11}. Additional details on the origins of this higher-order correction factor are given in Ref. \cite{AB11}. The dependence of the cross section on the square of the gluon distribution implies that exclusive vector meson production is a highly sensitive probe of gluon modifications in nuclei. Higher order corrections tend to decrease the rapidity distributions at mid-rapidity. The UPC photo-production cross section for the MSTW08LO distribution far exceeds the others and is multiplied by a factor 1/3 to better facilitate visualization in Figure \ref{f1}. It is also narrower than the photo-production cross sections obtained with the other PDFs. The CJ12 distribution is also much smaller than for the CT14 set.  At this point, we were not able to include the effects of errors in the PDFs  on the photo-production cross sections, which might help us understand the origins of these differences. Also, the MSTW08LO distribution is used as our standard set for the free proton PDF while the  CT14 set is taken as our standard of the higher-order corrections in the nuclear parton distributions. The large spread of the results of the values of the rapidity distributions and total cross sections is due to a lack of data in the low-x region producing large uncertainties in the parton distributions. Our results for the MSTW08 data set are somewhat different than those reported in Refs. \cite{AB11,AB12,ABM12} due to the different treatment we have introduced to calculate the virtual photon numbers.

\begin{figure}
\begin{center}
\includegraphics[scale=0.15]{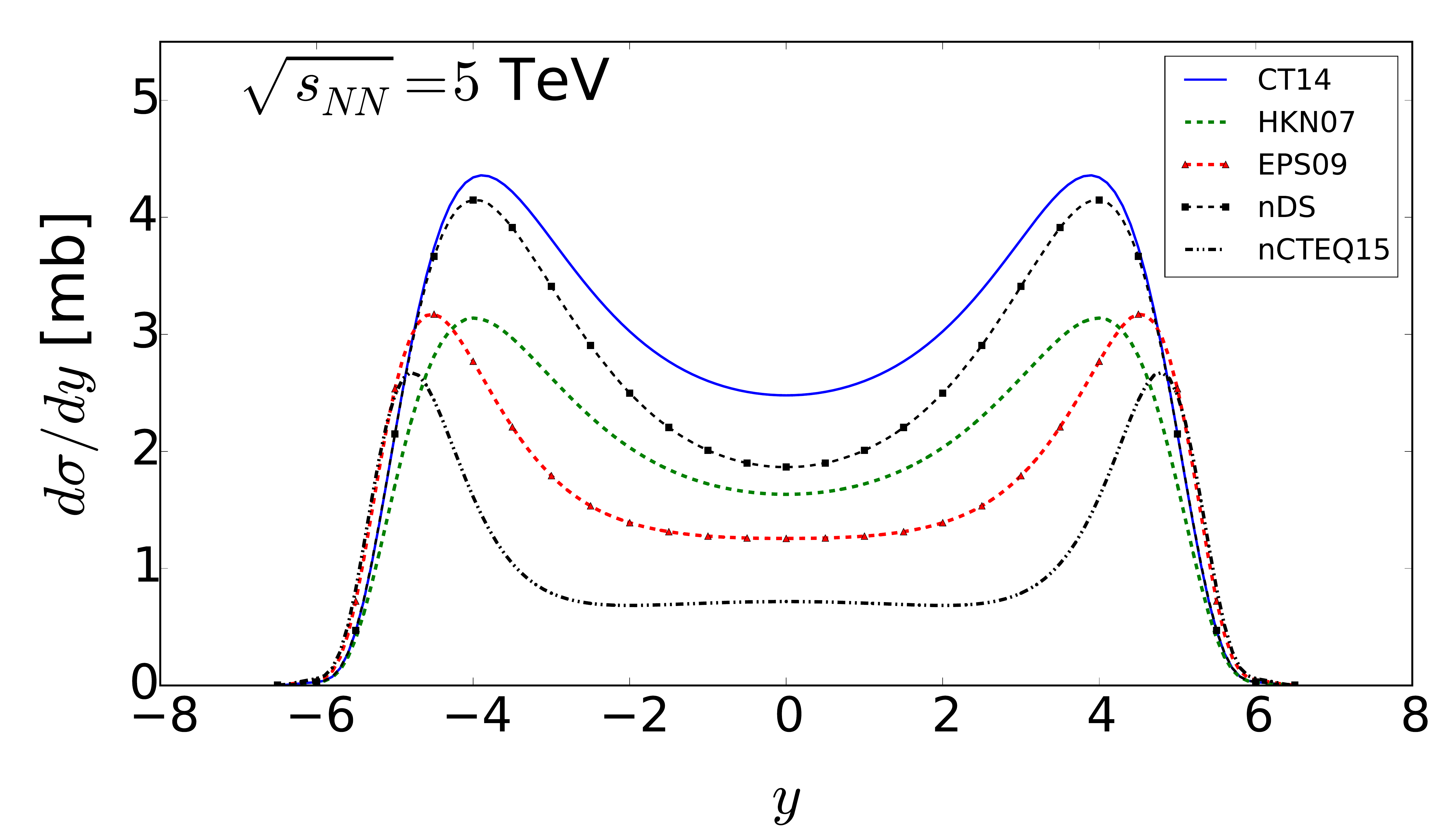}
\caption{Rapidity distributions (in mb) for J/$\psi$ production in UPCs at the LHC with $\sqrt{s_{NN}}=5$ TeV   using the EPS09, nDS and HKN07 medium corrections on  CT14 distributions are shown. Also shown are the results using nCTEQ15 distributions.} \label{f3}
\end{center}
\end{figure}

The NLO corrections modify the results with the MSTW08 distributions dramatically (by a factor of 1/3.5), but still do not bring it in line with the other distributions which also include higher-order effects. As discussed in the previous section,  free nucleon PDFs include valence and sea quarks which are probed by electrons, e.g., in HERA. As is evident from Eq. \eqref{gamA}, the UPCs probe the gluon distribution component of the PDFs through the gluon-fusion mechanism, which might lead to free $q\bar q$-pairs and bound states in the form of vector mesons. It is thus expected that high-order corrections, which include an increasing number of gluon-loop diagrams, would have a more discriminative effect of the photon-gluon fusion mechanism in UPCs. An obvious problem for interpretation of the calculations is that the rapidity distributions are obtained by an integration over the momentum fraction $x$ and it is thus not so straightforward to judge the several modifications of the PDFs in different regions of the $x$ space.  We also note that the factorization scale of $M_V/2 \sim 1 $ GeV lies in the lower bottom of all PDFs considered here, which have $Q^2$ values running from this energy and up.  

In the lower part of Figure \ref{f1} we show the results for  J/$\psi$  production at  $\sqrt{s_{NN}}=5$ TeV in Pb+Pb collisions included in Run 2 at the LHC. This is relevant for predictive purposes. Two obvious outcomes are observed: (a) the increase in the total cross sections, and (b) the wider rapidity distributions. The virtual photon spectrum now contains photons with higher energies, leading to a larger probability for the production of the J/$\psi$. They are also more abundantly produced closer to the beam rapidities and flatten out at mid-rapidity. At mid-rapidities ($y = 0$), the contribution from gluon shadowing is largest and it decreases so that around $y = 3$ it is overtaken by  anti-shadowing effects. At even higher  rapidities $y > 5$ anti-shadowing also loses its importance and the distribution is dominated by EMC and Fermi motion effects. This correlation between the rapidity distribution and the gluon distribution effects in the momentum fraction space $x$ was also reported in Ref. \cite{AB11}.  The regions of influence of shadowing and anti-shadowing effects are highlighted in the upper part of Figure \ref{f1}. It becomes evident that this process is an excellent probe of shadowing effects in the gluon distribution functions. Since the process depends quadratically on the gluon distributions the  rapidity distributions and total   J/$\psi$ production cross sections is strongly suppressed by shadowing effects even for the weak shadowing  present in the HKN07 case.

In Figure \ref{f2} we show our results for the rapidity distributions of J/$\psi$ production in UPCs at the LHC for $\sqrt{s_{NN}}=2.76$ TeV  using the EPS09, nDS and HKN07 medium corrections on  the CT14 distributions. Also shown are the results using the newest nCTEQ15 distributions. Here our calculations have been modified so that the upper figure displays only the coherent part J/$\psi$ production in UPCs at the LHC using the MSTW08 parton distribution function. The data are from Ref. \cite{Abb13}. In this case, and only for the MSTW08 set, we have constrained our calculation to coherent  photoproduction, i.e., when the photon acts on the whole nucleus. This effectively modifies the integrals in Eqs. (1-8) to J/$\psi$ by contributions satisfying the condition that transverse momenta be of the order of a few hundreds of MeV/c.  With such a restriction, the photon does not lead to nuclear breakup when the J/$\psi$ is produced. For larger transverse momenta the photon couples to a single nucleon and nuclear breakup can occur. The separation between coherent and incoherent photoproduction is introduced here in order to compare to the experiments which have been set to avoid incoherent photoproduction, characterized by large transverse momentum transfers \cite{Abb13}. 

The definition of coherent and incoherent collisions adopted in Ref. \cite{Abb13} assumes that the limiting transverse momentum is $p_T < 200$ MeV/c for coherent process and $p_T > 200$ MeV/c for incoherent processes, based in detection cuts introduced in the J/$\psi \rightarrow \mu^+\mu^-$ decay channel.  In in the J/$\psi \rightarrow e^+e^-$ decay channel, experimentalists have adopted the limiting value of $p_T < 300$ MeV/c  and $p_T > 300$ MeV/c,  respectively. For more details, see Ref. \cite{Abb13}. Theoretically, the separation of coherent and incoherent processes based on $p_T$ cuts can be carried out more effectively using the intranuclear cascade model of Ref. \cite{And15} which allows for the determination of the full four-momentum of the produced particle. Based on this model, the rapidity distribution for coherent processes using the MSTW08 distribution indicates that the form of the rapidity distribution does not change appreciably, except that their magnitude is effectively reduced by a factor of 0.5 \cite{Evan16}. Using this reduction factor in our calculations, we get the solid curve in bottom panel of Figure \ref{f2} for the MSTW08 distribution. The experimental data are well reproduced with the MSTW08 set with the inclusion of the EPS09 medium corrections. At mid-rapidity, the experimental data in Ref. \cite{Abb13} yields a combined weighted average for the di-muon and di-electron decay channels of $d\sigma^{coherent}/dy = 2.38^{+0.34}_{-0.24}$ mb, whereas our calculation yields 2.42 mb. Our results also agree with those obtained in Refs. \cite{AB11,AB12} for the same distribution set and medium corrections, if the reduction factor of 0.5 is also applied to their calculations. Due to a stronger gluon shadowing in the EPS08, it appreciably underestimates the rapidity distribution and cross sections by about a factor of two. The HKN07 and nDS parametrizations seem to contain too little gluon shadowing. As mentioned in Ref. \cite{Abb13}, PDF sets which do not include  gluon shadowing are incompatible with the measured results for J/$\psi$ vector meson production in UPCs at the LHC.

It is worthwhile mentioning at this point that, except for the results presented in the bottom panel of Figure \ref{f2},  all other results of our calculation are displayed without the separation between coherent and incoherent processes, i.e., both processes are weighted in with our integrals running over all possible momenta. As a last comment on the results presented in Figure \ref{f2} we notice that, being wider than the MSTW08 distributions (also evident in Figure \ref{f1}), the CJ12 and CT14 would only marginally reproduce the experimental data of \cite{Abb13} at the rapidities $y\sim -3$ and $y \sim 3$.  

\begin{table}
\begin{center}
\begin{tabular}
[c]{|l|l|l|l|}\hline
PDF              & $2.76$ TeV & $5$ TeV  \\\hline
MSTW08           & $83$       & $220$ \\
+ EPS08          & $7.8$      & $16$  \\
+ EPS09          & $42$       & $113$ \\
+ nDS            & $62$       & $155$ \\ \hline
CT14             & $26$       & $34$  \\
+ HKN07          & $19$       & $24$  \\
+ EPS09          & $16$       & $21$  \\
+ nDS            & $24$       & $30$  \\ \hline
CJ12             & $3.5$      & $4.5$ \\
+ HKN 07         & $2.5$      & $3.2$ \\
+ EPS 09         & $2.2$      & $2.8$ \\
+ nDS            & $3.1$      & $4.0$ \\ \hline
nCTEQ15          & $10.2$     & $13$  \\ \hline
\end{tabular}
\caption{Cross sections  for J/$\psi$  production in UPCs at the LHC for two laboratory energies and several parton distributions. Medium corrections  are included. \label{tab1}}
\end{center}
\end{table}

In Figure \ref{f3} we show the rapidity distributions for J/$\psi$ production in UPCs at the LHC with $\sqrt{s_{NN}}=5$ TeV   using the EPS09, nDS and HKN07 medium corrections on   CT14 distributions. Also shown are the results using nCTEQ15 distributions. We observe an increase in the total cross sections, and  wider rapidity distributions due to the increase in the collision energy and the presence of harder photons in the equivalent photon spectrum. In contrast to the MSTW08 distributions all others PDFs consistently yield similar widths for $d\sigma/dy$, as also observed in Figures \ref{f1} and \ref{f2}.  However, they display different shadowing effects, as emphasized by the spread of about a factor of 2 of $d\sigma/dy$ at mid-rapidities.

In Table \ref{tab1} we present the cross sections  for J/$\psi$  production in UPCs at the LHC for two laboratory energies and several parton distributions. Results obtained with medium corrections  are included separately. We observe sizable differences among the different PDF sets. The effects of shadowing in the gluon distribution functions  and their contribution to the total photo-production cross sections are the same as those discussed for the case of the rapidity distributions. As shadowing  effects are responsible for reducing the rapidity distributions in the range $-3 < y<3$, where $d\sigma/dy$ are largest, the total cross sections are also affected accordingly.

\begin{table}
\begin{center}
\begin{tabular}
[c]{|l|l|l|l|}\hline
PDF        & $2.76$ TeV & $5$ TeV  \\\hline
MSTW08     & $17$       & $40$    \\
+ EPS08    & $3.5$      & $6.7$   \\
+ EPS09    & $7.6$      & $17$    \\
+ nDS      & $15$       & $34$    \\ \hline
CT14       & $8.0$      & $11.3$  \\
+ HKN07    & $6.0$      & $8.4$   \\
+ EPS09    & $5.3$      & $7.3$   \\
+ nDS      & $7.1$      & $9.8$   \\ \hline
CJ12       & $1.1$      & $1.6$   \\
+ HKN07    & $0.81$     & $1.2$   \\
+ EPS09    & $0.72$     & $1.0$   \\
+ nDS      & $1.0$      & $1.4$   \\ \hline
nCTEQ15      & $3.9$      & $5.5$   \\ \hline
\end{tabular}
\caption{Cross sections (in mb)  for $\psi$(2S)  production in UPCs at the LHC for two laboratory energies and several parton distributions. Medium corrections  are included. \label{tab2}}
\end{center}
\end{table}

\begin{figure}
\begin{center}
\includegraphics[scale=0.41]{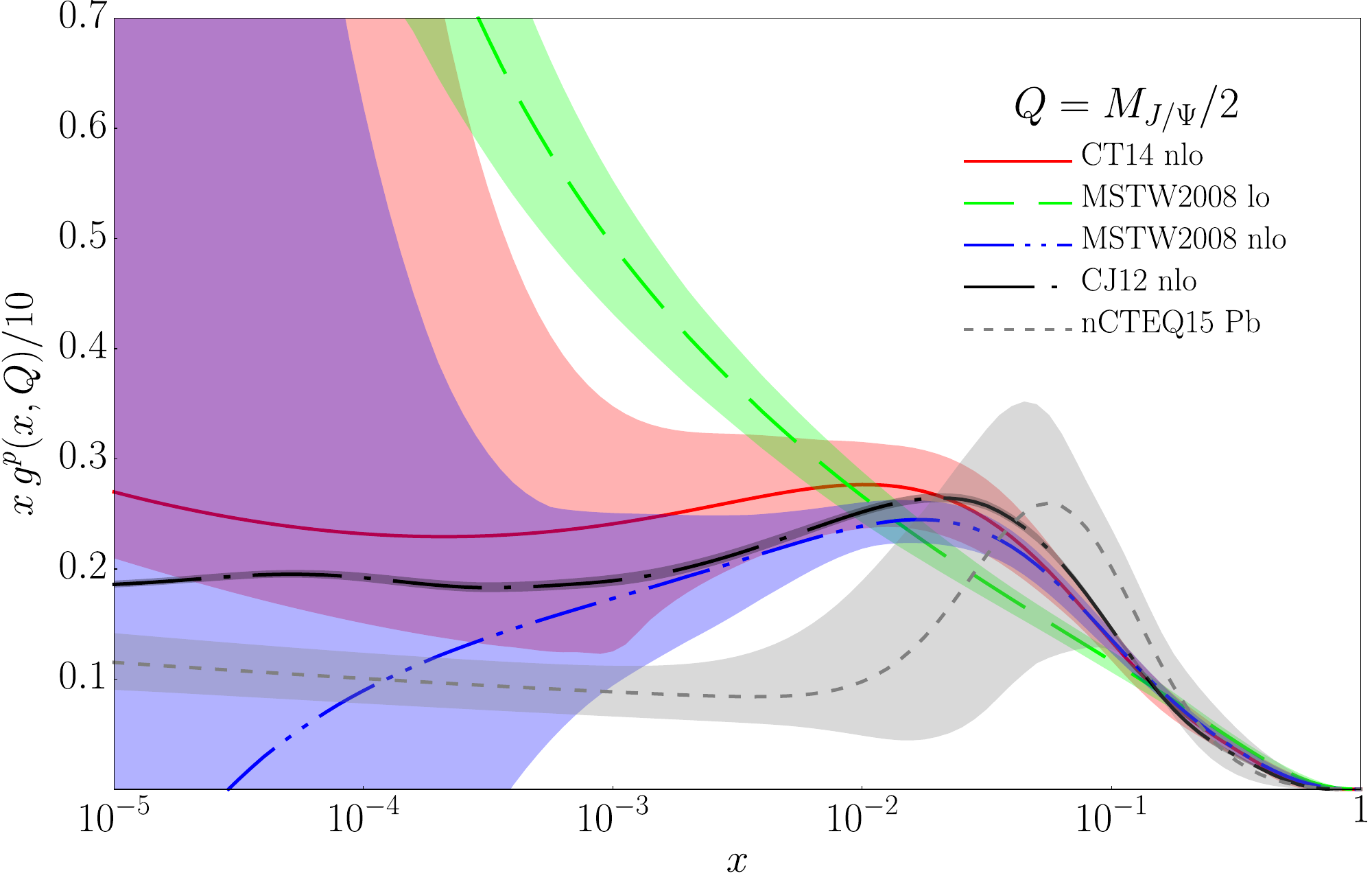}
\includegraphics[scale=0.43]{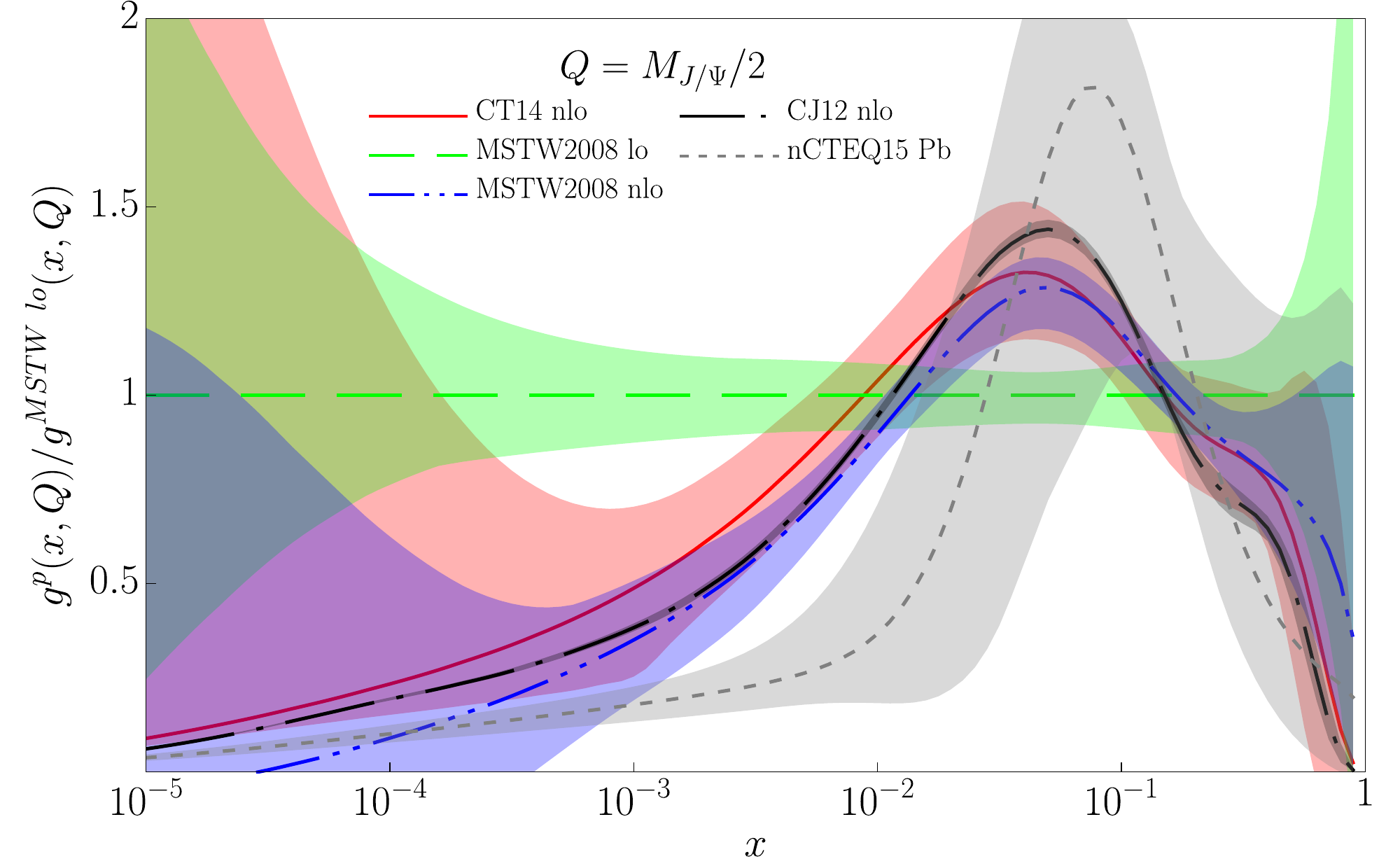}
\caption{The longitudinal momentum $x$ times the PDF for the gluon with errors  at $Q = M_{J/\psi}/2$ for MSTW08 \cite{MSTW09}, CJ12 \cite{Owe13}, CT14 \cite{Say14},  and nCTEQ15 Lead \cite{GN14} PDF releases (upper figure) and their ratios over the  MSTW08 LO set (lower figure). Differences in the proton PDFs contributing to the differences in  cr4oss sections are clearly visable. The nCTEQ15 contains additional medium effects  which contribute to an enhancement of the PDF in the high-$x$ region.}\label{pdfs}
\end{center}
\end{figure} 

It is worthwhile at this point to discuss a severe problem with the determination of the most appropriate PDF set to reproduce data from UPCs at the LHC. To make this point explicit, we show in Figure \ref{pdfs} the longitudinal momentum $x$ times the PDF for the gluon with errors  at $Q = M_{J/\psi}/2$ for MSTW08 \cite{MSTW09}, CJ12 \cite{Owe13}, CT14 \cite{Say14},  and nCTEQ15 Lead \cite{GN14} PDF releases (upper figure) and their ratios over the  MSTW08 LO set (lower figure). Differences in the proton PDFs contributing to the differences in  cross sections are clearly visable. The nCTEQ15 contains additional medium effects  which contribute to an enhancement of the PDF in the high-$x$ region. In the present work we have used the average $x$-dependence of the several PDF sets without taking into account the errors in the individual sets. As we see in Figure \ref{pdfs}, the errors are quite large in the region probed by $J/\psi$ production in UPCs at the LHC. Had we included these errors, there would be a substantial overlap of the results presented in Figures \ref{f1}-\ref{f3} and also in the following figures for $\psi(2S)$ and $\Upsilon$ production. In particular, it would be hard to determine the best PDF set reproducing  the data from Ref. \cite{Abb13}, as shown in Figure \ref{f2}.

\subsection{$\psi(2S)$ production in UPCs at $\sqrt{s_{NN}}=2.76$ TeV and $\sqrt{s_{NN}}=5$ TeV }

\begin{figure}
\begin{center}
\includegraphics[scale=0.16]{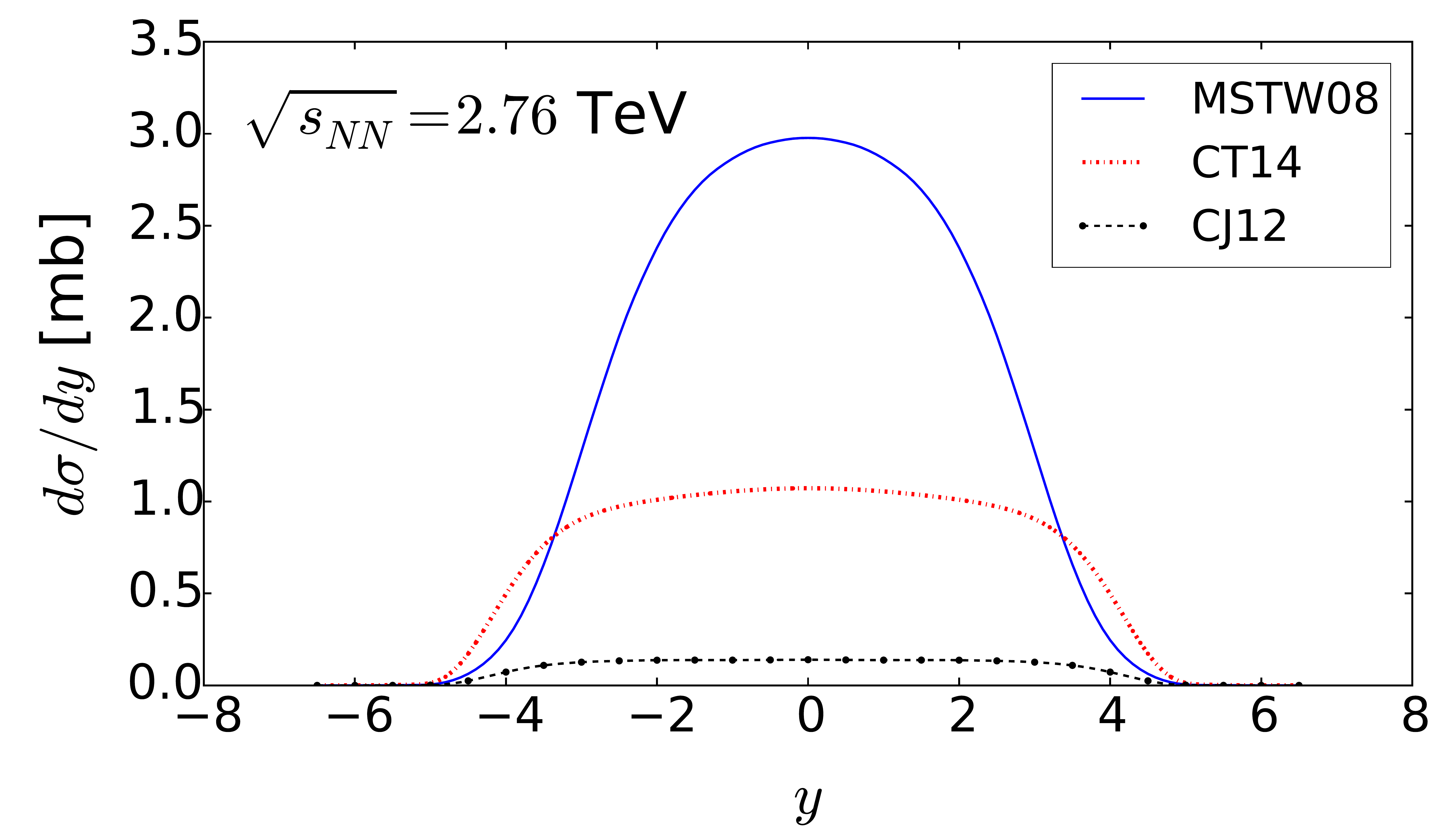}
\includegraphics[scale=0.16]{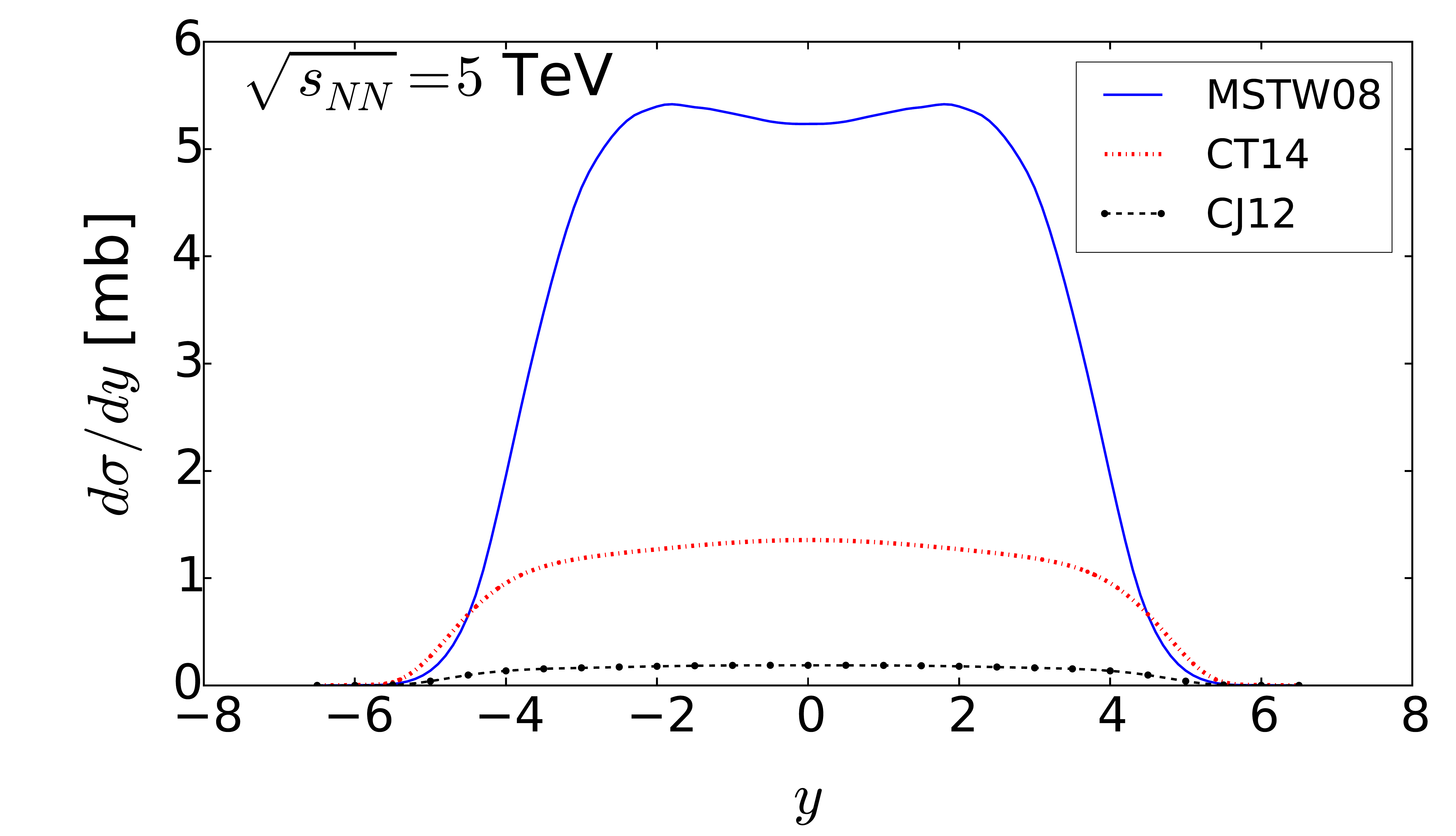}
\caption{Rapidity distributions (in mb) for $\psi(2S)$ production in UPCs at the LHC with $\sqrt{s_{NN}}=2.76$ TeV (upper figure) and  $\sqrt{s_{NN}}=5$ TeV (lower figure) using the MSTW08 \cite{MSTW09},  CJ12 \cite{Owe13}, and CT14 \cite{Say14} PDF releases. The contribution of the MSTW08 set in LO includes a correction factor $\zeta_V$ to accommodate NLO corrections.} \label{f4}
\end{center}
\end{figure} 

Here we present our calculations for the production of $\psi(2S)$ UPCs at the LHC for $\sqrt{s_{NN}}=2.76$ TeV and $\sqrt{s_{NN}}=5$ TeV. Although its mass is only 20\% larger than the J/$\psi$ mass ($M_{\psi(2S)} = 3.686$ GeV), the results reported for  $\psi(2S)$ production are different from those for J/$\psi$ production.  

In Figure \ref{f4} we show the rapidity distributions (in mb) for $\psi(2S)$ production in UPCs at the LHC with $\sqrt{s_{NN}}=2.76$ TeV (upper figure) and  $\sqrt{s_{NN}}=5$ TeV (lower figure) using the PDF releases MSTW08 \cite{MSTW09},  CJ12 \cite{Owe13}, and CT14 \cite{Say14}. We recall that the contribution of the MSTW08 in this case, and also for the production of $\Upsilon$'s, does not include a correction factor $\zeta_V$ to accommodate NLO corrections, as it was done for the J/$\psi$.  The $\psi(2S)$ and $\Upsilon$ photoproduction cross sections, seem to be best reproduced without this correction. 

\begin{figure}
\begin{center}
\includegraphics[scale=0.16]{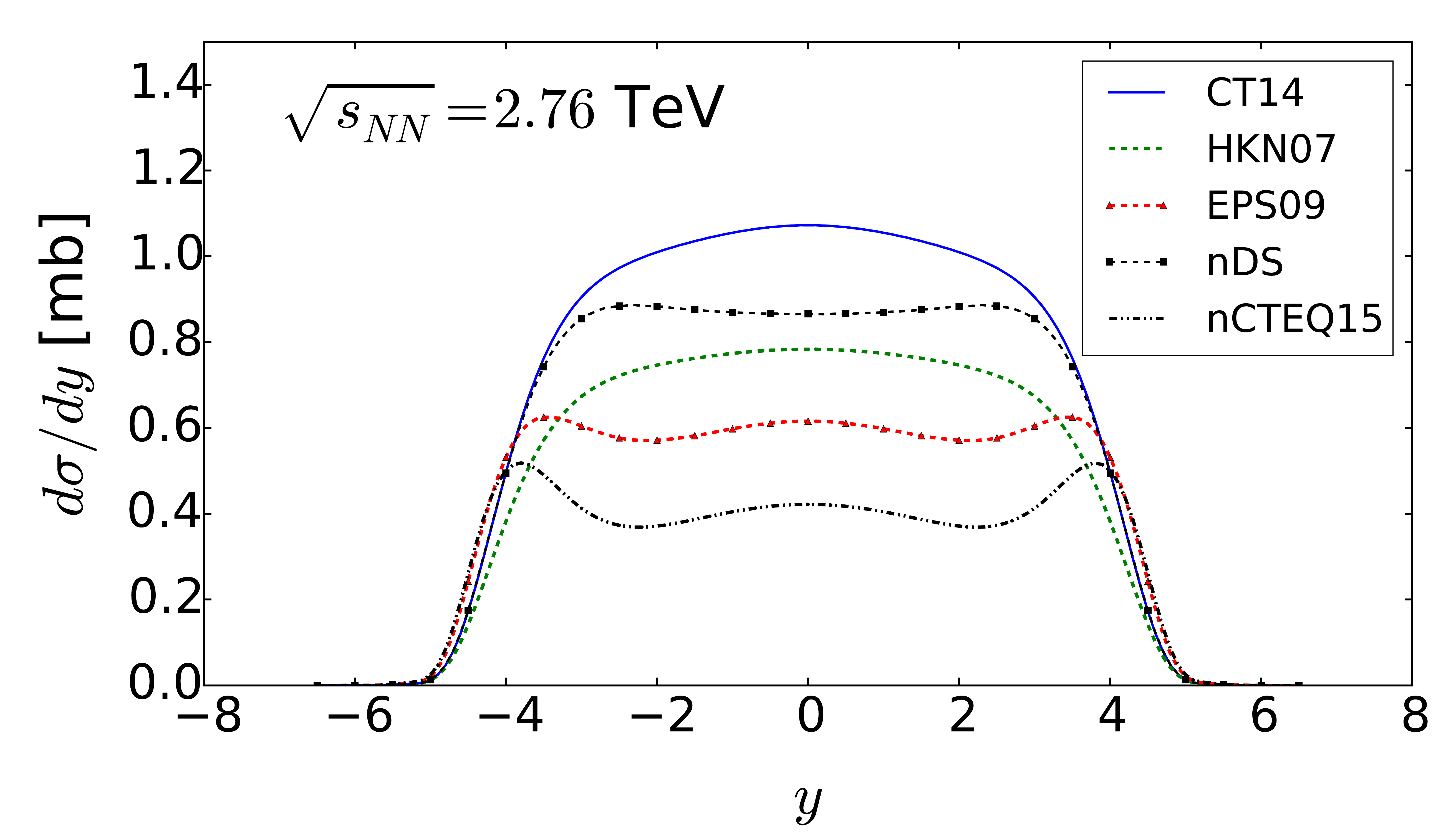}
\caption{Rapidity distributions (in mb) for $\psi(2S)$ production in UPCs at the LHC with $\sqrt{s_{NN}}=2.76$ TeV   using the EPS09, nDS and HKN07 medium corrections on the  CT14 release. Also shown are the results using nCTEQ15 distributions. } \label{f5}
\end{center}
\end{figure}

\begin{figure}
\begin{center}
\includegraphics[scale=0.16]{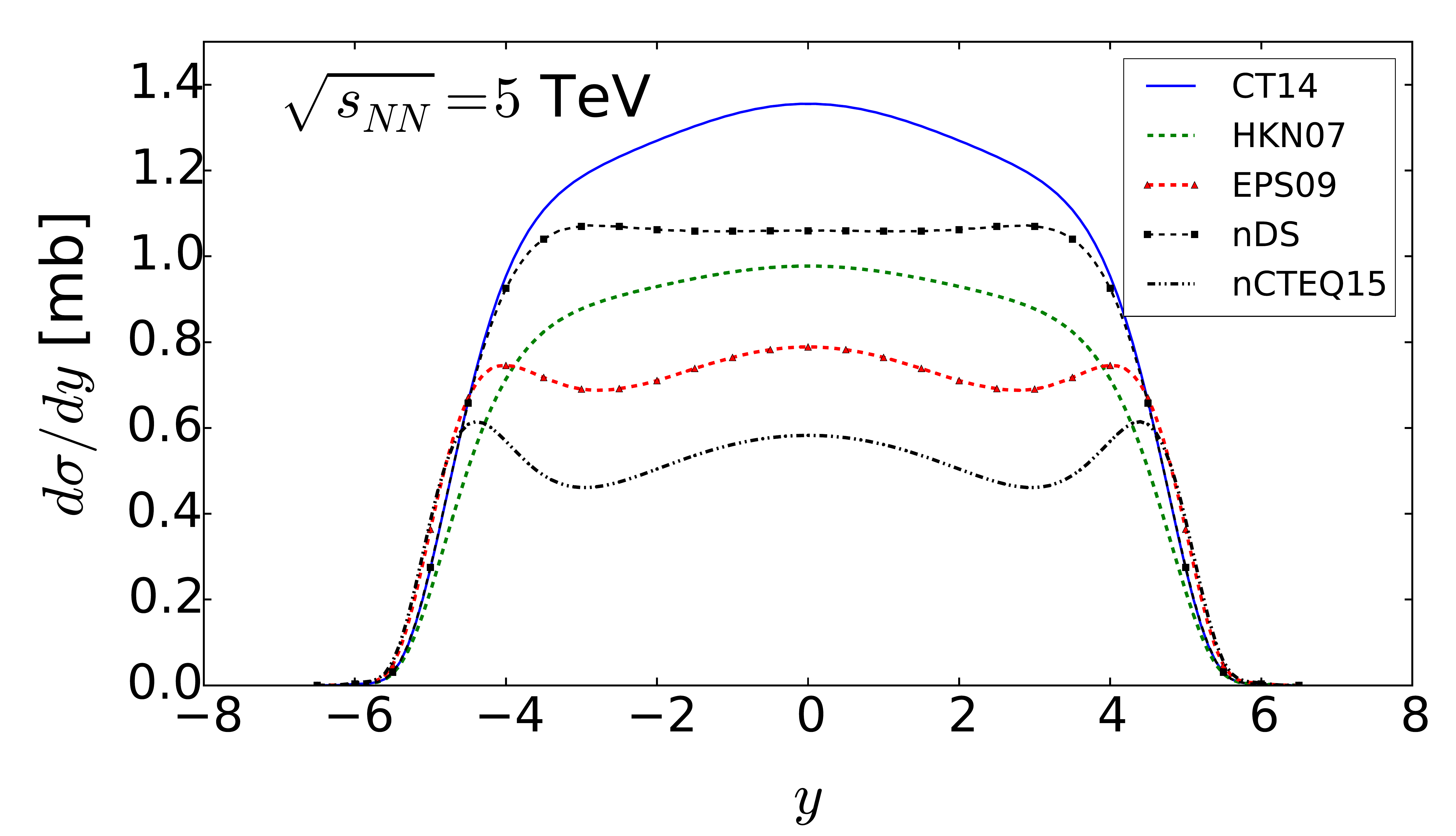}
\caption{Rapidity distributions (in mb) for $\psi(2S)$ production in UPCs at the LHC with $\sqrt{s_{NN}}=5$ TeV   using the EPS09, nDS and HKN07 medium corrections on  CT14  release. Also shown are the results using nCTEQ15 distributions. } \label{f6}
\end{center}
\end{figure}

In Figure \ref{f5} the rapidity distributions for $\sqrt{s_{NN}}=2.76$ TeV are obtained using the EPS09, nDS and HKN07 medium corrections on CT14 distributions. Also shown are the results using nCTEQ15 distributions. In Figure \ref{f6} the rapidity distributions shown are calculated for $\sqrt{s_{NN}}=5$ TeV  using the EPS09, nDS and HKN07 medium corrections on  CT14  distributions. Also shown are the results using nCTEQ15 distributions.

Besides the increase of the mass leading to a reduction of the magnitude of the cross sections, also due the smaller magnitude of the leptonic decay width, $\Gamma_{ee}=2.37$ keV for the $\psi(2S)$, there is an evident spread of the cross sections for the different parton distributions.   This can be seen in Table \ref{tab2} where we present our cross sections  for $\psi(2S)$  production in UPCs at the LHC for two laboratory energies and several parton distributions. Results obtained with medium corrections  are included separately. We observe some differences among the different PDF sets, but mainly in the magnitude of the cross sections.  

The variations of our results for  the several PDFs and medium corrections shown in Figures \ref{f4}-\ref{f6} and in Table \ref{tab2} are due to the sensitivity of the calculations on the meson mass in the integration over the different regions of the parton momentum fraction $x$.  As an example, the sensitivity (or lack thereof) of the cross sections to medium corrections can be inferred  by using the free MSTW08 and the medium-corrected MSTW08+nDS. For the J/$\psi$ production at $\sqrt{s_{NN}}=2.76$ TeV and $\sqrt{s_{NN}}=5$ TeV, the reduction due to shadowing effects are 13\% and 30\%  respectively, while for  $\psi(2S)$ production the reductions amount to 12\% and 15\%. Therefore, at least the total cross sections seem to  globally probe shadowing and anti-shadowing effects in a distinguishable way for   J/$\psi$ and $\psi(2S)$  at higher bombarding energies.

The different regions of the momentum fraction, $x$,  probed in $\psi(2S)$ production in UPCs at the LHC are more evident if we consider the plots for the rapidity distributions, shown in Figures \ref{f4}-\ref{f6}. It is evident that the main changes occur at the tail of the distributions where the passage from shadowing to anti-shadowing should occur, specially at  $\sqrt{s_{NN}}=2.76$ TeV. As the beam energy increases to $\sqrt{s_{NN}}=5$ TeV the width of the distributions become very similar to those displayed earlier for  J/$\psi$ production. This applies to medium corrected PDFs as well as those for free protons.

\subsection{$\Upsilon$ production in UPCs at $\sqrt{s_{NN}}=2.76$ TeV and $\sqrt{s_{NN}}=5$ TeV }

\begin{figure}
\begin{center}
\includegraphics[scale=0.16]{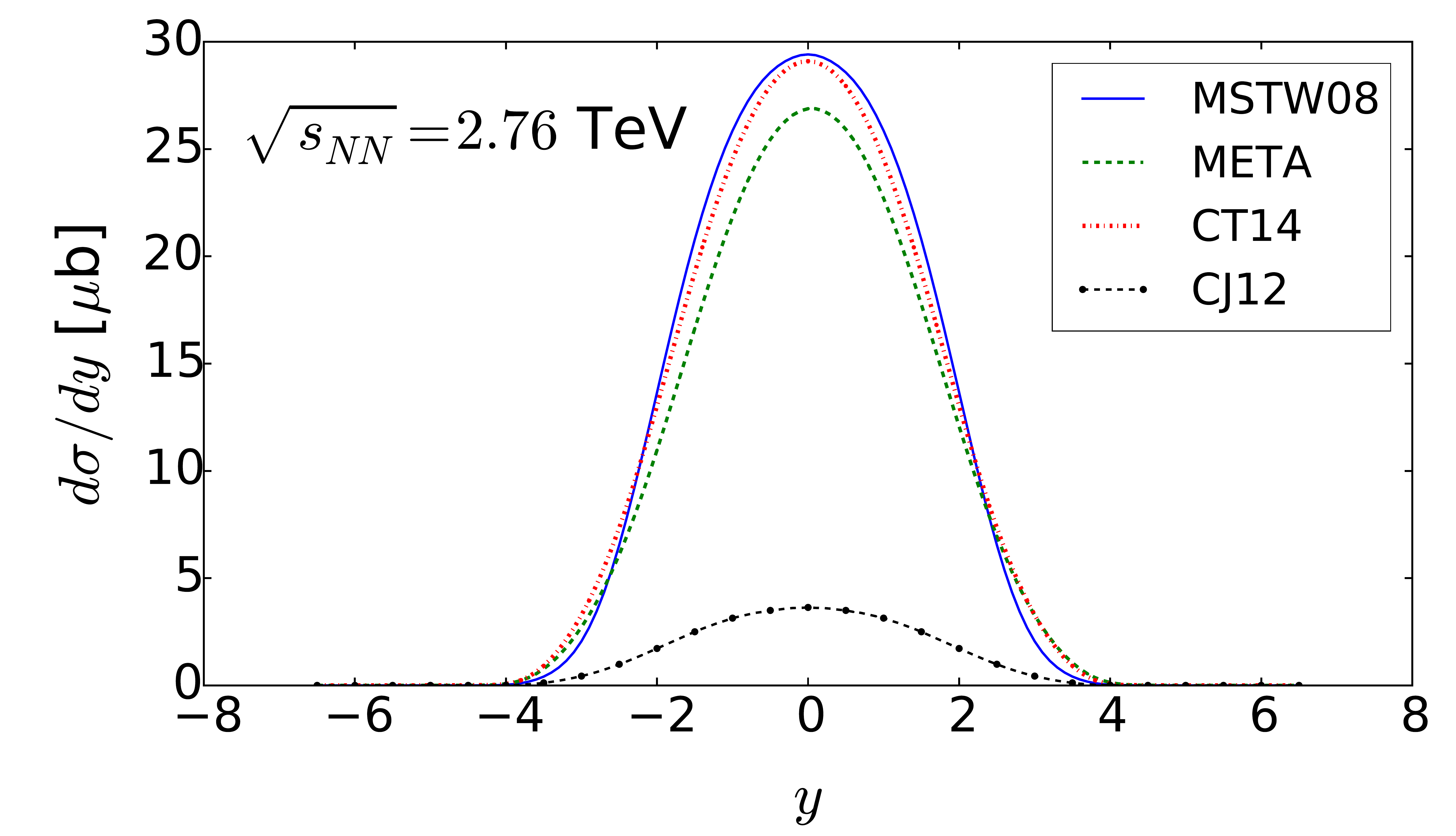}
\includegraphics[scale=0.16]{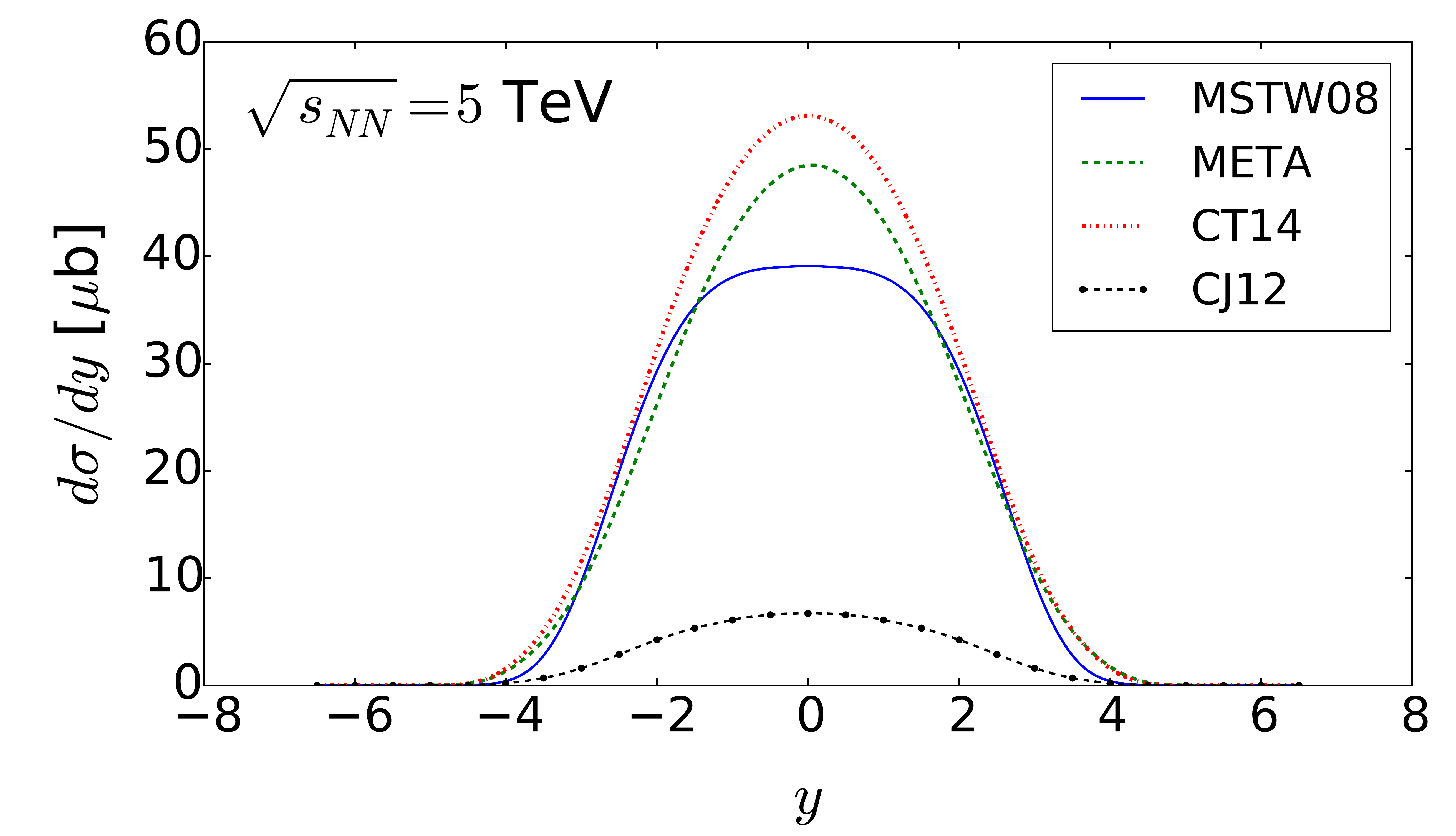}
\caption{Rapidity distributions for $\Upsilon$ production in UPCs at the LHC with $\sqrt{s_{NN}}=2.76$ TeV (upper figure) and  $\sqrt{s_{NN}}=5$ TeV (lower figure) using the PDF compilations MSTW08 \cite{MSTW09},  CJ12 \cite{Owe13}, CT14 \cite{Say14}, and META14 \cite{GN14}. The contribution of the MSTW08 set in LO includes a correction factor $\zeta_V$.} \label{f7}
\end{center}
\end{figure}

\begin{figure}
\begin{center}
\includegraphics[scale=0.16]{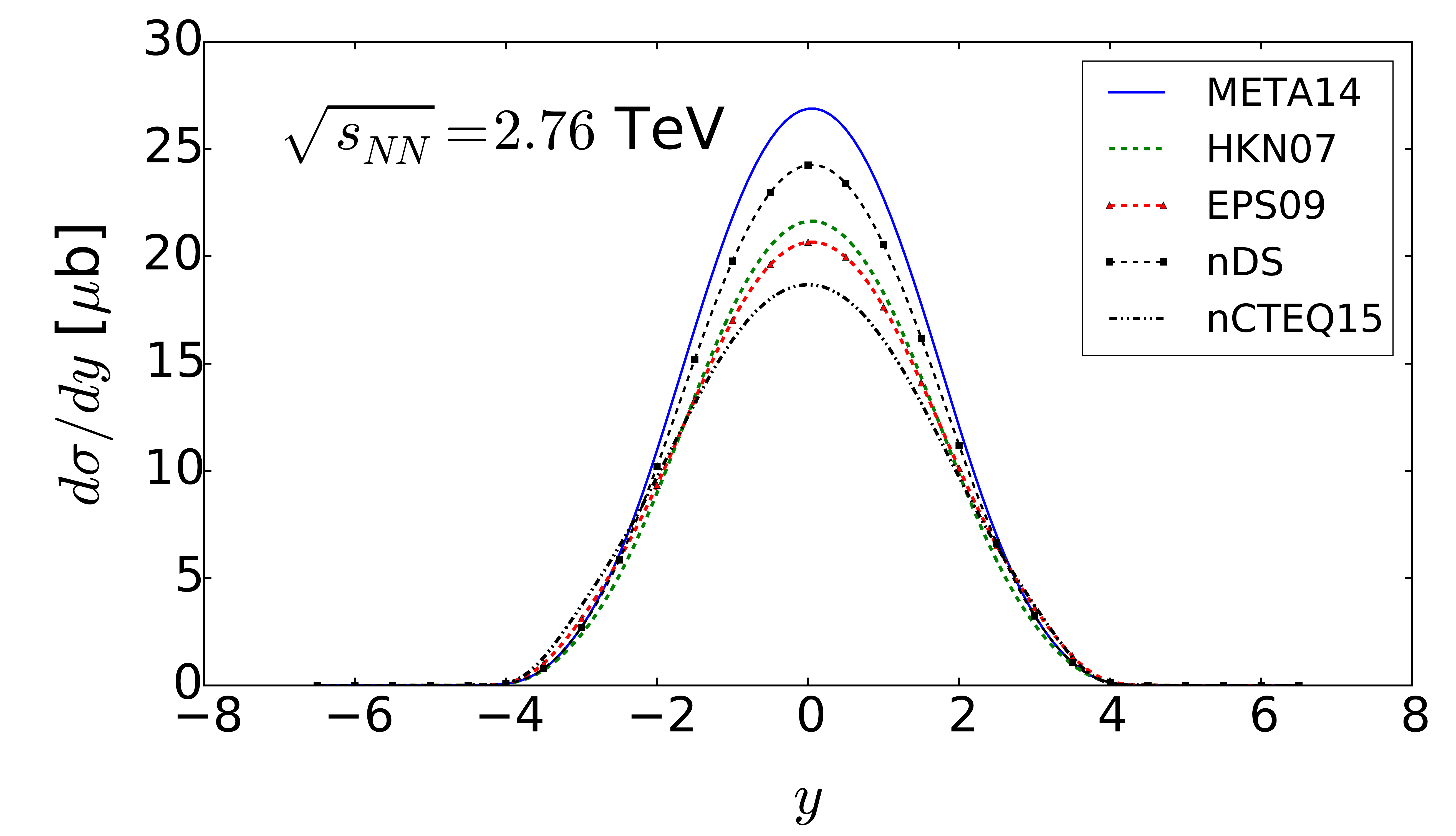}
\includegraphics[scale=0.16]{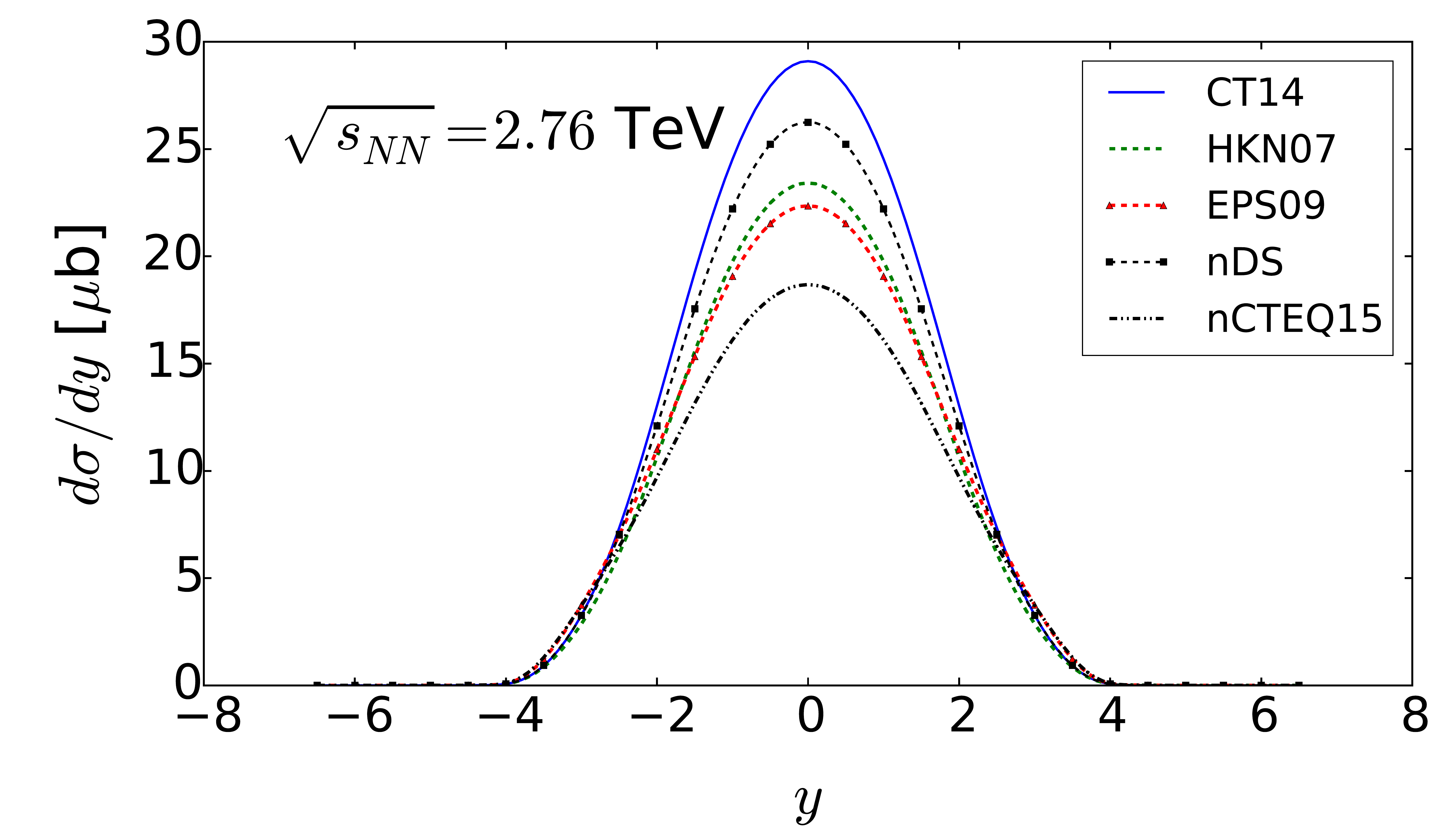}
\caption{Rapidity distributions (in $\mu$b) for $\Upsilon$ production in UPCs at the LHC with $\sqrt{s_{NN}}=2.76$ TeV   using the EPS09, nDS and HKN07 medium corrections on  META14  distributions (upper figure) and CT14 (lower figure). Also shown are the results using nCTEQ15 distributions. } \label{f8}
\end{center}
\end{figure}

\begin{figure}
\begin{center}
\includegraphics[scale=0.16]{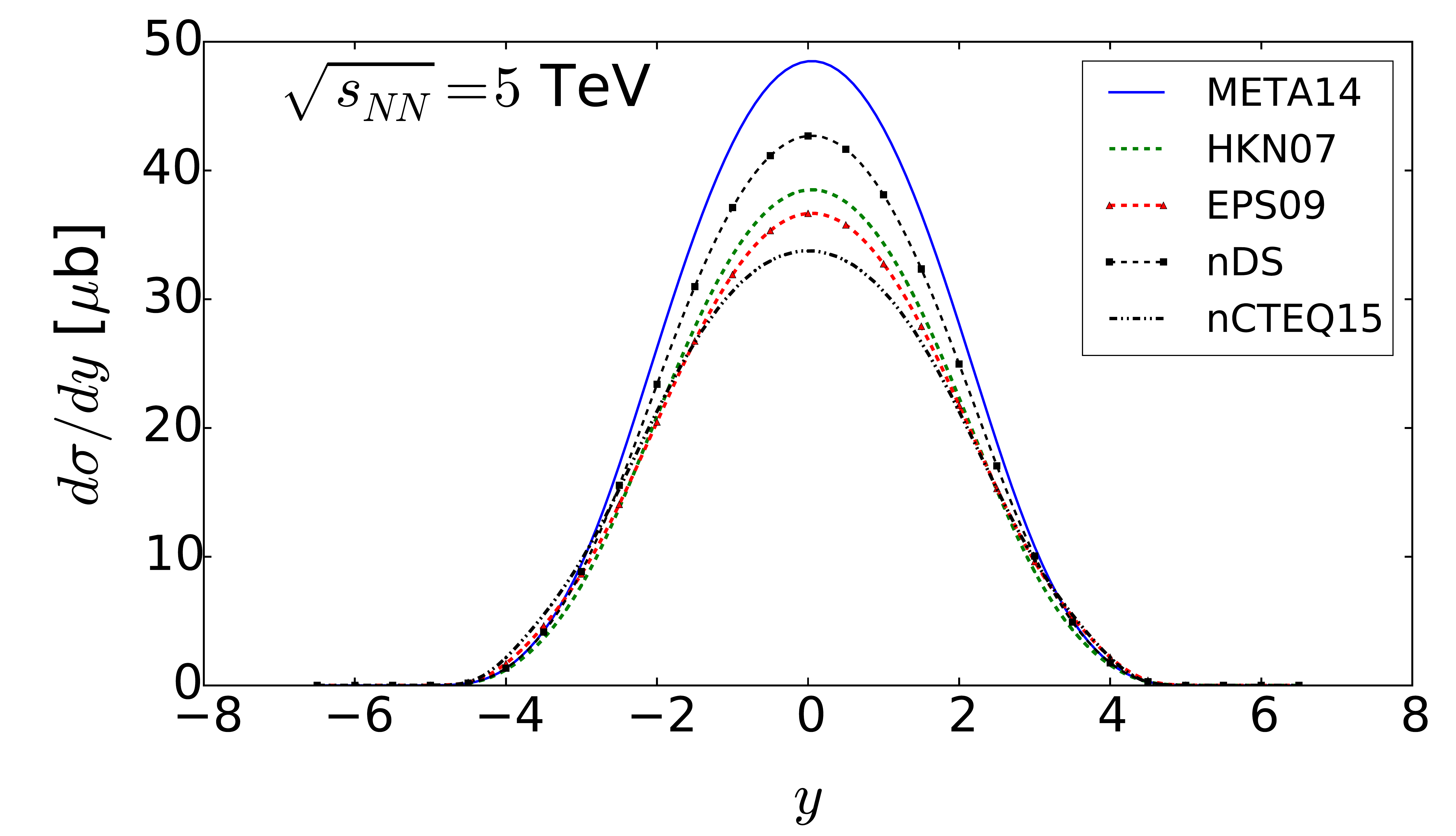}
\includegraphics[scale=0.16]{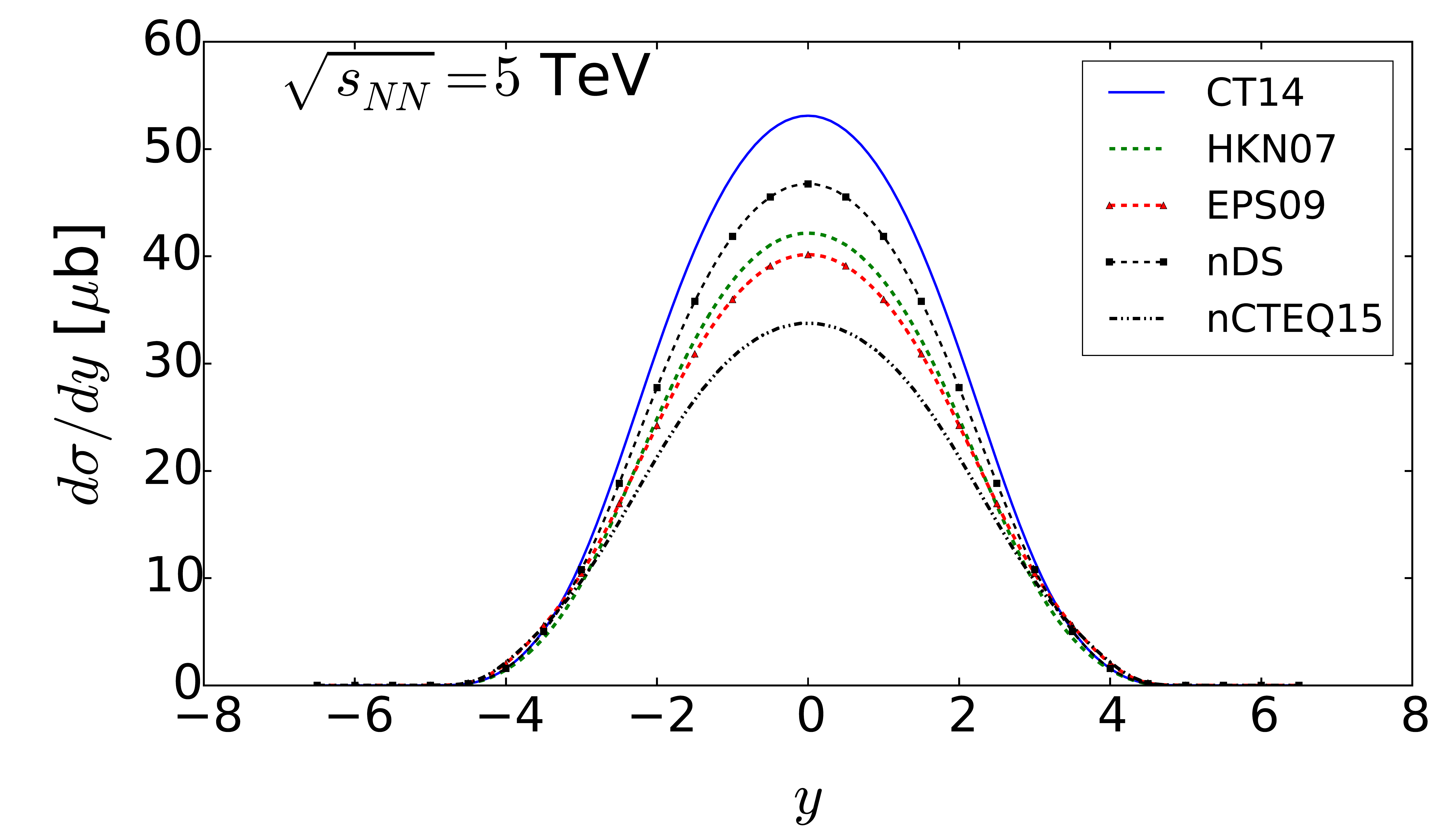}
\caption{Rapidity distributions (in $\mu$b)  for $\Upsilon$ production in UPCs at the LHC with $\sqrt{s_{NN}}=5$ TeV   using the EPS09, nDS and HKN07 medium corrections on  META14  distributions (upper figure) and CT14 (lower figure). Also shown are the results using nCTEQ15 distributions. } \label{f9}
\end{center}
\end{figure}

Here we present our calculations for the production of $\Upsilon(1S)$ in UPCs at the LHC for $\sqrt{s_{NN}}=2.76$ TeV and $\sqrt{s_{NN}}=5$ TeV. In this case, the mass is a factor of 3 larger than the J/$\psi$ mass ($M_{\Upsilon} = 9.46$ GeV), and the total cross sections decrease dramatically, by 3 orders of magnitude due to the much smaller number of quasi-real photons in this energy region.  The rapidity distributions and cross sections (in $\mu$b)  for $\Upsilon$  production in UPCs at the LHC for two laboratory energies and several parton distributions using the leptonic decay width, $\Gamma_{ee}=1.34$ keV, are shown in Figures \ref{f7}-\ref{f9} and Table \ref{tab3}. Medium corrections  are included separately. 

In Figure \ref{f7} we plot the rapidity distributions for $\Upsilon$ production in UPCs at the LHC with $\sqrt{s_{NN}}=2.76$ TeV (upper figure) and  $\sqrt{s_{NN}}=5$ TeV (lower figure) using the PDF compilations MSTW08 \cite{MSTW09},  CJ12 \cite{Owe13}, CT14 \cite{Say14}, and META14 \cite{GN14}. Here it is appropriate to include the META14 PDF, because of the large $\Upsilon$ mass. 

In Figure \ref{f8} we show our results for  the EPS09, nDS and HKN07 medium corrections on  META14  distributions (upper figure) and CT14 (lower figure) for $\sqrt{s_{NN}}=5$ TeV. Also shown are the results using nCTEQ15 distributions. 

Finally, in Figure \ref{f9}, we plot the rapidity distributions  using the EPS09, nDS and HKN07 medium corrections on  META14  distributions (upper figure) and CT14 (lower figure) for $\sqrt{s_{NN}}=5$ TeV. Also shown are the results using nCTEQ15 distributions. 

It is noticeable in Figures \ref{f7}-\ref{f9} that the larger mass of the $\Upsilon$ has as the overall effect of reducing the width of the rapidity distributions with the contribution from shadowing effects being limited to  $-2 < y < 2$. As we noticed for $\psi(2S)$ production, the  impact of the  larger mass of the $\Upsilon$ modifies the momentum fraction $x$ content probed by the square of the gluon distribution, through the $d\sigma/dy$ dependence on $[xg_A(x)]^2$, leading to a reduction of the shadowing effects probed in the rapidity scale.  But, most importantly, the larger value of the $\xi$ variable in the virtual photon spectrum leads to a much larger reduction of the cross sections.  The relative size of $\sigma$ and $d\sigma/dy$ obtained with each PDF vary appreciably for the three mesons and also for different energies. Thus, a better scrutiny of parton distribution functions could be achieved in future experiments by measuring UPCs.     

\begin{table}
\begin{center}
\begin{tabular}
[c]{|l|l|l|l|}\hline
PDF        & $2.76$ TeV & $5$ TeV  \\\hline
MSTW08     & $112$      & $194$  \\
+ EPS08    & $60$       & $88$   \\
+ EPS09    & $78$       & $124$  \\
+ nDS      & $105$      & $176$  \\ \hline
CT14       & $111$      & $237$  \\
+ HKN07    & $90$       & $189$  \\
+ EPS09    & $90$       & $186$  \\
+ nDS      & $101$      & $211$  \\ \hline
CJ12       & $14$       & $31$   \\
+ HKN07    & $12$       & $25$   \\
+ EPS09    & $11$       & $16$   \\
+ nDS      & $13$       & $24$   \\ \hline
nCTEQ15      & $78$       & $161$   \\ \hline
\end{tabular}
\caption{Cross sections (in $\mu$b)  for $\Upsilon(1S)$  production in UPCs at the LHC for two laboratory energies and several parton distributions. Medium corrections  are included separately. \label{tab3}}
\end{center}
\end{table}

\section{Conclusions}

We have studied the elastic photoproduciton of the vector mesons J/$\psi$, $\psi(2S)$ and  $\Upsilon(1S)$ in ultra-peripheral collisions at the Large Hadron Collider at CERN. We have focused on Pb+Pb collisions at two energies: $\sqrt{s_{NN}}=2.76$ TeV and  $\sqrt{s_{NN}}=5$ TeV. This study is complimentary to those exploring PDFs to predict the outcome of particle production in central collisions between relativistic heavy ion collisions. In this work we have only considered direct photo-production in UPCs, although it is known that the resolved photon components also play a sizable role \cite{AB12}.  The EPS09 model, which includes a coherent partonic energy loss, still seems to be the best fit to the experimentally available data on J/$\psi$ production \cite{Abb13}. We have  also utilized a larger number of parton distribution functions than have been considered so far, such as the META, CT14, nDS and nCTEQ15 PDFs, with or without medium corrections.

The total cross sections as well as the rapidity distributions are appreciably sensitive to shadowing at mid-rapidities and a rather insensitive to antishadowing at very forward and very backward rapidities. Because of their small difference in mass there are not many noticeable differences in their sensitivity to gluon distributions for the  cross sections and rapidity distributions in J/$\psi$ and $\psi(2S)$ production. But there is a drop of the cross section by a factor of 3 or larger, due to the reduction in the gluon PDF and the reduced number of virtual photons for the larger vector meson mass.  The situation changes dramatically for the case of $\Upsilon$ production because the cross sections decrease by 3 orders of magnitude to a much larger vector meson mass. But also the rapidity distributions become much narrower in this case. 

The quadratic dependence of the elastic photoproduction on gluon distributions and their medium modifications  do allow for a serious probe for the different parton distribution functions. The discrimination of shadowing effects can manifest through cross sections differing by factors  as large as  10 for the different PDFs we have considered. We hope that the Run 2 of PbPb collisions  at the LHC will allow to better constrain  medium modifications of gluon distribution and the consequent magnitude of shadowing effects.

Theoretical models still lack the inclusion of final state breakup of vector mesons, which seems to be evidenced in Ref. \cite{Abb13}. As claimed in Ref. \cite{And15}, where cascade simulations have been carried out, only a small fraction of the produced J/$\psi$ leave the nucleus, emphasizing the  importance of final state interactions in vector meson production in UPCs. Such effects would be better investigated if additional nuclei could be used in the collider. It is also worthwhile mentioning that the role of resolved photon partonic distributions still demands additional theoretical work. Finally, the PDF sets display appreciable uncertainties as a function of the momentum fraction $x$. The experimental data on UPCs at the LHC for rapidity distributions and total cross sections of vector meson production introduces an additional tool to constrain PDF distributions, although an assessment of the role of the errors in the $x$-dependence of the PDFs deserves a better scrutiny. Work in this direction is in progress.

\section*{Acknowledgements} 

This work was supported in part by the U.S. DOE grants DE-FG02-08ER41533 and  the U.S. NSF Grant No. 1415656. This work was also partially supported by the U.S. Department of Energy under grant DE-FG02- 13ER41996.



\begin{thebibliography}{99}
\bibitem{Zepto} Gian Francesco Giudice, ``A Zeptospace Odyssey: A Journey into the Physics of the LHC", Oxford University Press, Oxford (2010).

\bibitem{HE13} E.M. Henley,  S.D. Ellis,  eds.,  ``100 Years of Subatomic Physics", World Scientific (2013).

\bibitem{Eng64} F. Englert and R. Brout, Phys. Rev. Lett. 13, 321 (1964).

\bibitem{Hig64} P.  Higgs,  Phys. Rev. Lett. 13, 508 (1964).

\bibitem{Gur64}  G. Guralnik, C.R.  Hagen, and T.W.B.  Kibble, Phys. Rev. Lett. 13, 585 (1964).

\bibitem{Miy66} H. Miyazawa, Prog. Theor. Phys. 36, 1266 (1966).

\bibitem{DG81} S. Dimopoulos and H. Georgi, Nucl. Phys. B 193, 150 (1981).

\bibitem{Mur00} H. Murayama,  ``Supersymmetry phenomenology", arXiv:hep-ph/0002232  (2000).

\bibitem{Kal21} T. Kaluza,  Preuss. Akad. Wiss. Berlin. (Math. Phys.), p. 966972 (1921).

\bibitem{Kle26} O. Klein,  Zeit. Phys. A 37, 895 (1926).

\bibitem{RS99} L. Randall, R. Sundrum,  Phys. Rev. Let. 83, 3370 (1999).

\bibitem{Ran02} Lisa Randall, ``Extra Dimensions and Warped Geometries", Science 296, 5572 (2002).

\bibitem{CP75} J. C. Collins and M. J. Perry, Phys. Rev. Lett. 34, 1353 (1975).

\bibitem{HN77} H. Bohr, and H.B. Nielsen,  Nucl. Phys. B 128,  275 (1977).

\bibitem{Zaj08} W.A. Zajc, Nucl. Phys. A 805, 283c (2008).

\bibitem{Mat86} T. Matsui and H. Satz,  Phys. Lett. B 178 (1986).

\bibitem{Al05} B. Alessandro et al.,  Eur. Phys. J. C 39, 335 (2005).

\bibitem{Abe09} B.I. Abelev et al.,  Phys. Rev., C80, 41902 (2009).

\bibitem{Amal07} R. Arnaldi et al.,  Phys. Rev. Lett. 99, 132302 (2007); Nucl. Phys. A 830, 345c  (2009).

\bibitem{ATL11} ATLAS collaboration, Phys. Lett. B 697, 294 (2011).

\bibitem{Ada11}A. Adare et al., Phys. Rev. C 84, 054912 (2011).

\bibitem{ALI12} ALICE collaboration, Phys. Rev. Lett. 109, 072301 (2012).

\bibitem{CMS12} CMS collaboration, J. High Energy Phys. 05, 063 (2012).

\bibitem{BB94} C.A. Bertulani and G. Baur,  ``Relativistic heavy ion physics without nuclear contact", Physics Today, March, p. 22 (1994). 

\bibitem{BB88} C.A. Bertulani and G. Baur, Phys. Reports 163,  299 (1988).

\bibitem{Rit93} J. Ritman, et al., Phys. Rev. Lett. 70, 533 (1993).

\bibitem{Sch93} R. Schmidt, et al., Phys. Rev. Lett. 70, 1767 (1993).

\bibitem{Bau96} G. Baur, et al., Phys. Lett. B368, 251 (1996).

\bibitem{Bla98} G. Blanford, et al., Phys. Rev. Lett. 80, 3040 (1998).

\bibitem{GB02} V. P. Goncalves and C. A. Bertulani, Phys. Rev. C 65, 054905 (2002).

\bibitem{Fran02} L. Frankfurt, M. Strikman, and M. Zhalov, Phys. Lett. B 540, 220 (2002).

\bibitem{Gut13} V. Guzey, E. Kryshen, M. Strikman, and M. Zhalov, Phys. Lett. B 726, 290 (2013).

\bibitem{AB11} A. Adeluyi and C. A. Bertulani, Phys. Rev. C 84, 024916 (2011).

\bibitem{AB12} A. Adeluyi and C. A. Bertulani, Phys. Rev. C 85, 044904 (2012).

\bibitem{AN13} A. Adeluyi and T. Nguyen, Phys. Rev. C 87, 027901 (2013).

\bibitem{Reb12} V. Rebyakova, M. Strikman, and M. Zhalov, Phys. Lett. B 710, 647 (2012).

\bibitem{CMS14} CMS Collaboration, Report Number CMS-PAS-HIN-12-009 (2014).

\bibitem{Abb13} E. Abbas et al., Eur. Phys. J. C 73,  2617 (2013)

\bibitem{Abe13} B. Abelev et al., Phys. Lett. B 718, 1273 (2013).

\bibitem{Abe14} B. Abelev et al., J. High Energy Phys. 02, 073 (2014).

\bibitem{Ada15} J. Adam et al., Phys. Lett. B 751, 358 (2015).

\bibitem{GM05} V. P. Goncalves and M. V. T. Machado, Eur. Phys. J. C 40, 519 (2005).

\bibitem{GM06} V. P. Goncalves and M. V. T. Machado, Phys. Rev. C 73, 044902 (2006); Phys. Rev. D 77, 014037 (2008); Phys. Rev. C 80, 054901 (2009).

\bibitem{SS07} W. Schafer and A. Szczurek, Phys. Rev. D 76, 094014 (2007).

\bibitem{RS08} A. Rybarska, W. Schafer and A. Szczurek, Phys. Lett. B 668, 126 (2008).

\bibitem{AG08} A. L. Ayala Filho, V. P. Goncalves and M. T. Griep, Phys. Rev. C 78, 044904 (2008).

\bibitem{Bert09} C.A. Bertulani, Phys. Rev. C 79, 047901 (2009).

\bibitem{CE10} C.A. Bertulani and M. Ellermann, Phys. Rev. C 81, 044910 (2010).

\bibitem{GM11} V. P. Goncalves and M. V. T. Machado, Phys. Rev. C 84, 011902 (2011).

\bibitem{CS12} A. Cisek, W. Schafer and A. Szczurek, Phys. Rev. C 86, 014905 (2012).

\bibitem{GM12} V. P. Goncalves and M. M. Machado, Eur. Phys. J. C 72, 2231 (2012)

\bibitem{GM14}  V. P. Goncalves and M. M. Machado, Eur. Phys. J. A 50, 72 (2014).

\bibitem{GMN14} V. P. Goncalves, B. D. Moreira and F. S. Navarra, Phys. Rev. C 90, 015203 (2014).

\bibitem{GMN15} V. P. Goncalves, B. D. Moreira and F. S. Navarra, Phys. Lett. B 742, 172 (2015).

\bibitem{Bau02} G. Baur, Kai Hencken, D. Trautmann, S. Sadovsky, Y. Kharlov, Phys. Reports 364, 359 (2002).

\bibitem{BKN05} C. A. Bertulani, S. R. Klein, and J. Nystrand, Annu. Rev. Nucl. Part. Sci. 55, 271 (2005).

\bibitem{Bal08} A. Baltz et al., Phys. Rep. 458, 1 (2008).

\bibitem{And15} E. Andrade-II and I. Gonzalez and A. Deppman and C. A. Bertulani, Phys. Rev. C 92, 064903 (2015).

\bibitem{ABM12} Adeola Adeluyi, C. A. Bertulani, and M. J. Murray, Phys. Rev. C 86, 047901 (2012).

\bibitem{BFF90}  G. Baur and L. G. Ferreira Filho, Nucl. Phys. A  518, 786 (1990).

\bibitem{CJ90} R. N. Cahn and J. D. Jackson, Phys. Rev. D42, 3690 (1990).

\bibitem{GRV92} M. Gluck, E. Reya, and A. Vogt, Phys. Rev. D 46, 1973 (1992).

\bibitem{SS95} G. A. Schuler and T. Sjostrand, Z. Phys. C 68, 607 (1995).

\bibitem{Cor04} F. Cornet, P. Jankowski, and M. Krawczyk, Acta Phys. Pol. B 35, 2215 (2004).

\bibitem{KNV02} S. R. Klein, J. Nystrand, and R. Vogt, Phys. Rev. C 66, 044906 (2002).

\bibitem{Rys93} M. G. Ryskin, Z. Phys. C 57, 89  (1993).

\bibitem{Brod94} S. J. Brodsky, L. Frankfurt, J. F. Gunion, A. H. Mueller, and M. Strikman, Phys. Rev. D 50, 3134 (1994).

\bibitem{Rys97}  M. G. Ryskin, R. G. Roberts, A. D. Martin, and E. M. Levin, Z. Phys. C 76, 231  (1997). 

\bibitem{FKS98} L. Frankfurt, W. Koepf, and M. Strikman, Phys. Rev. D 57, 512 (1998).
 

\bibitem{FMS01} L. Frankfurt, M. McDermott, and M. Strikman, J. High Energy Phys. 03, 045  (2001). 

\bibitem{Sus00} K. Susuki, A. Hayashigaki, K. Itakura, J. Alam, and T. Hatsuda, Phys. Rev. D 62, 031501 (2000).

\bibitem{Ji04} X. Ji, Annu. Rev. Nucl. Part. Sci.  54, 413 (2004).

\bibitem{Collins:1987pm}
   J.~C.~Collins and D.~E.~Soper,
      Ann.\ Rev.\ Nucl.\ Part.\ Sci.\  {\bf 37}, 383 (1987).
   doi:10.1146/annurev.ns.37.120187.002123
      

\bibitem{GW73} D.J. Gross, F. Wilczek, Phys. Rev. Lett. 30, 1343 (1973); Phys. Rev. D8,  3633 (1973).

\bibitem{Pol73} H. D. Politzer,  Phys. Rev. Lett. 30, 1346 (1973);  Phys. Rep. 14, 129 (1974).

\bibitem{Pum02} J. Pumplin, D. R. Stump, J. Huston, H. L. Lai, P. M. Nadolsky, and W. K. Tung, J. High Energy Phys. 07, 012 (2002).

\bibitem{KLO04} S. Kretzer, H. Lai, F. Olness and W. Tung, Phys, Rev. D 69,  114005 (2004). 

\bibitem{Nad08} P.M. Nadolsky et al., Phys. Rev. D78, 013004 (2008).

\bibitem{MSTW09} A.D. Martin, W.J. Stirling, R.S. Thorne and G.Watt, Eur. Phys. J. C63, 189 (2009).

\bibitem{GJR08} M. Gl\"uck, P. Jimenez-Delgado and E. Reya, Eur. Phys. J. C 53, 355 (2008).

\bibitem{Bal10} R. D. Ball, Nucl. Phys. B 838, 136 (2010).

\bibitem{ABKM10} S. Alekhin, J. Bl\"umlein, S. Klein and S. Moch, Phys. Rev. D81, 014032 (2010).

\bibitem{Aa10} F. D. Aaron et al., ``H1 and ZEUS collaborations", J. High Energy Phys. 01, 109 (2010).

\bibitem{Arm06} N. Armesto, J. Phys. G 32, R367 (2006).

\bibitem{EKS99} K. J. Eskola, V. J. Kolhinen, and C. A. Salgado, Eur. Phys. J. C 9, 61 (1999).

\bibitem{FS04} D. de Florian and R. Sassot, Phys. Rev. D 69, 074028 (2004).

\bibitem{HKS04} M. Hirai, S. Kumano, and T. H. Nagai, Phys. Rev. C 70, 044905 (2004)

\bibitem{HKS05} M. Hirai, S. Kumano, and T. H. Nagai, Nucl. Phys. Proc. Suppl. 139, 21 (2005).

\bibitem{HKN07} M. Hirai, S. Kumano, and T. H. Nagai, Phys. Rev. C 76, 065207 (2007).

\bibitem{EPS08} K. J. Eskola, H. Paukkunen, and C. A. Salgado, J. High Energy Phys. 07, 102 (2008).

\bibitem{EPS09} K. J. Eskola, H. Paukkunen, and C. A. Salgado, J. High Energy Phys. 04, 065 (2009).

\bibitem{FGS05} L. Frankfurt, V. Guzey, and M. Strikman, Phys. Rev. D 71, 054001 (2005).

\bibitem{GR78} M. Gl\"uck and E. Reya, Phys. Lett. 79B, 453  (1978).

\bibitem{GRV95} M. Gl\"uck, E. Reya, and A. Vogt, Z. Phys. C 67, 433  (1995).

\bibitem{EKR98} K. J. Eskola, V. J. Kolhinen, and P. V. Ruuskanen, Nucl. Phys. B535, 351 (1998).

\bibitem{AG01} A. L. Ayala and V. P. Gon\c calves, Eur. Phys. J. C 20, 343 (2001).

\bibitem{Aub83} J. J. Aubert et al. (European Muon Collaboration), Phys. Lett. B 123, 275 (1983).

\bibitem{GST95} D. F. Geesaman, K. Saito, and A.W. Thomas, Annu. Rev. Nucl. Part. Sci. 45, 337 (1995).

\bibitem{PW00} G. Piller and W. Weise, Phys. Rep. 330, 1 (2000).

\bibitem{Jag74} C. W. De Jager, H. De Vries, and C. De Vries, At. Data Nucl. Data Tables 14, 479 (1974).

\bibitem{PDG}  	K.A. Olive et al. (Particle Data Group), Chin. Phys. C 38, 090001 (2014).

\bibitem{Kov15} K. Kovarik, A. Kusina, T. Jezo, D. B. Clark, C. Keppel, F. Lyonnet, J. G. Morfin, F. I. Olness, J. F. Owens, I. Schienbein, J. Y. Yu, arXiv:1509.00792.

\bibitem{Owe13} J. F. Owens, A. Accardi, W. Melnitchouk, Phys. Rev. D 87, 09012 (2013).

\bibitem{Say14} Sayipjamal Dulat, Tie Jiun Hou, Jun Gao, Marco Guzzi, Joey Huston, Pavel Nadolsky, Jon Pumplin, Carl Schmidt, Daniel Stump, C. P. Yuan, arXiv:1506.07443.

\bibitem{GN14} Jun Gao, Pavel Nadolsky,  J. High Energy Phys. 35, 10.1007 (2014).

\bibitem{Evan16}  E. Andrade-II, private communication. 




\end{thebibliography}
\end{document}